\documentclass[superscriptaddress,twocolumn,showkeys,aps,prb,reprint]{revtex4-2}

\usepackage{graphicx}
\usepackage{dcolumn}
\usepackage{bm}
\usepackage{float}
\usepackage{color}
\usepackage{amsmath}
\usepackage{multirow}

\usepackage{calligra}
\usepackage[T1]{fontenc}

 \usepackage{nicefrac}
\usepackage{soul,xcolor}
\usepackage{latexsym}
\usepackage{amsthm}
\usepackage{bbm}
\usepackage{amsfonts}
\usepackage{amssymb}
\usepackage{hyperref}

\usepackage{wrapfig}

\hypersetup{colorlinks=true, citecolor=blue}
\setstcolor{red}

\newcommand{\pp}{\mathbf{p}}
\newcommand{\kk}{\mathbf{k}}
\newcommand{\rr}{\mathbf{r}}

\newcommand{\Hcal}{\mathcal{H}}

\newcommand{\Ccal}{\mathcal{C}}

\newcommand{\PMX}{P-$\mathrm{MX'}$}
\newcommand{\PXM}{P-$\mathrm{XM'}$}
\newcommand{\PXX}{P-$\mathrm{XX'}$}

\newcommand{\APH}{AP-$\mathrm{2H}$}

\newcommand{\uvec}[1]{\mathbf{\hat{#1}}}

\newcommand{\hly}{{\rm MX}_2}
\newcommand{\ely}{{\rm M'X'}_2}
\newcommand{\aaa}{\mathbf{a}}
\newcommand{\AAA}{\mathbf{A}}

\newcommand{\ket}[1]{| #1 \rangle}
\newcommand{\braoket}[3]{\langle #1 | #2 | #3 \rangle}

\allowdisplaybreaks

\begin{document}

\title{Floquet engineering of topological bands in semiconductor van der Waals heterobilayers}

\author{Er\'endira Santana-Su\'arez}
\affiliation{Instituto de F\'isica, Universidad Nacional Aut\'onoma de M\'exico, Ciudad de M\'exico, C.P. 04510, M\'exico}

\author{Brayan E.\ Walteros-Mendivelso}
\affiliation{Universidad Distrital Francisco Jos\'e de Caldas, Carrera 3 No.\ 26A-40, PCLF, 110311, Bogot\'a D.C., Colombia}

\author{A.\ Jazm\'in Tapia-de-la-Rosa}
\affiliation{Instituto de F\'isica, Universidad Nacional Aut\'onoma de M\'exico, Ciudad de M\'exico, C.P. 04510, M\'exico}

\author{Mahmoud M.\ Asmar}
\email{masmar@kennesaw.edu}
\affiliation{Department of Physics, Kennesaw State University, Marietta, Georgia, 30060, USA}

\author{David A.\ Ruiz-Tijerina}
\email{d.ruiz-tijerina@fisica.unam.mx}
\affiliation{Instituto de F\'isica, Universidad Nacional Aut\'onoma de M\'exico, Ciudad de M\'exico, C.P. 04510, M\'exico}

\date{\today}

\begin{abstract}
 
We show that periodic driving with near-{infrared} to visible light {drives} topological phase transitions in the {photon-dressed band structure} of transition-metal dichalcogenide heterobilayers. We apply the Floquet formalism to a {light-coupled} lowest-order $\kk\cdot\pp$  Hamiltonian, and obtain an effective four-band model that correctly captures the essential photon-dressed bands of type-II TMD heterobilayers {in the vicinity of the first photon resonance.} Crossings between the zero- and one-photon {sectors effectively invert the bands}, yielding topological phases with Chern numbers up to $\pm 2$ and gaps of order 10 meV. Our results establish TMDs as prime candidates for engineering bands with higher Chern numbers, and exploring {driven} topological states in solid state media.

\end{abstract}

\maketitle

\emph{Introduction.} The search for two-dimensional (2D) topological quantum materials has numerous motivations; from the development of more efficient electronic devices\cite{ComponentsZhang,TopoTransistorEzawa,TopoTransistorLi,TopoTransistorPeeters,TopoTransistorXing,TopoTransistorFuhrer} to the realization of exotic states of matter\cite{Exotic1,Exotic2,Exotic3,Exotic4,Exotic5}. In the latter case, single-particle bands with Chern numbers $|\Ccal|>1$ are of special interest as potential hosts of topologically nontrivial many-body states of a qualitatively different nature to Landau levels\cite{NoLLAnalogue1,NoLLAnalogue2,NoLLAnalogue3,NoLLAnalogue4,NoLLAnalogue5,NoLLAnalogue6,NoLLAnalogue7,NoLLAnalogue8}. A key challenge is identifying realistic platforms in which such high-Chern bands can be {robustly generated and controlled}. Amongst other platforms for topological matter, 2D materials stand out for the tunability of their electronic states under external electric\cite{Tunability1,Tunability2,Tunability3,Tunability4,Tunability5,Tunability6,Tunability7} and strain\cite{Strain1,Strain2,Strain3,Strain4,Strain5} fields, in principle allowing for controlled and reversible transitions into and out of topological phases through band inversions.

Semiconducting transition-metal dichalcogenides (TMDs) have long been regarded as potential topological playgrounds\cite{Xiao2012,Cazalilla2014,TatianaR2023,Tong2017,Wu2019,DasSarma2020}. {Though lacking a true topological band gap across the entire Brillouin zone (BZ)}, TMDs can exhibit the valley Hall effect\cite{MoS2valleyHall,2LMoS2valleyHall,IntrinsicValleyHallMoS2} when a valley population imbalance is induced, which highlights the nontrivial topology of the valley states in these systems. These states can be further exploited for band engineering through the formation of vertical heterostructures\cite{GeimGrigorieva}. For instance, Tong et al.\cite{Tong2017} envisioned the possibility of electrically-driven conduction-valence band inversion in type-II TMD heterobilayers with sufficiently small band gaps {to access quantum spin Hall instulator (QSHI) phases, though posing a considerable challenge to current experimental capabilities.}

\begin{figure}[t!]
    \centering
    \includegraphics[width=\columnwidth]{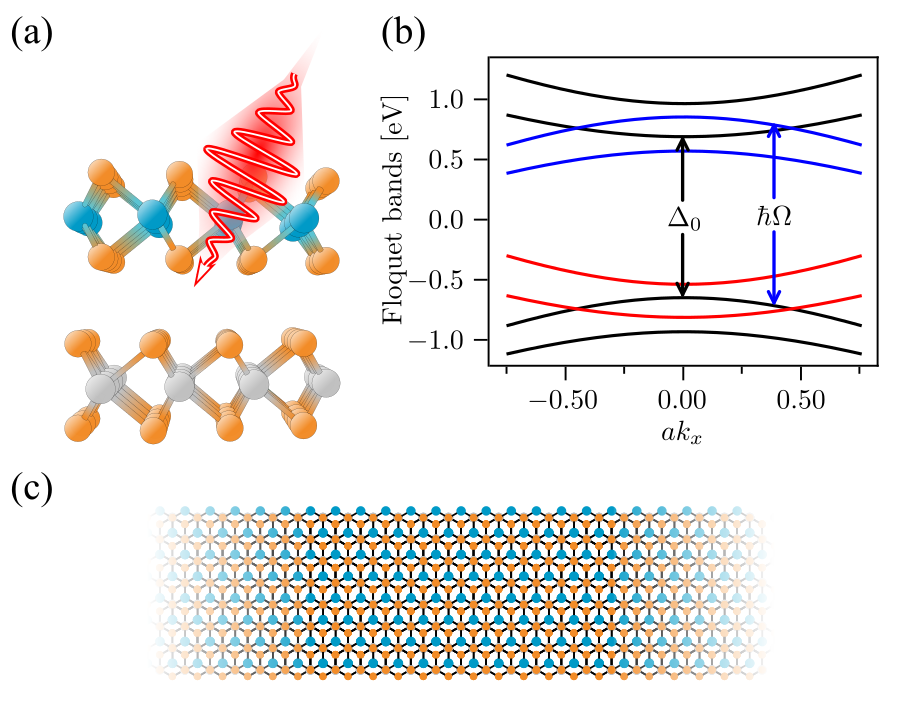}
    \caption{(a) Schematic of a near-IR laser pulse impinging on a P-stacked WSe${}_2$/MoSe${}_2$ hetero-bilayer. (b) Zero-field Floquet-Bloch bands for $\hbar\Omega \gtrsim \Delta_0$, in an extended Floquet BZ scheme. The equilibrium ($m=0$) bands are shown in black, whereas the $m=1$ ($m=-1$) replicas of the valence (conduction) band are shown in blue (red). Clear four-band sectors form, separated by energies of order $\Delta_0$. (c) Illustration of a \PMX\, zig-zag nanoribbon.}
    \label{fig:crossing}
\end{figure}

{Periodic driving provides an alternative route to topological band engineering that avoids some limitations of static setups}~\cite{flreview1,flreview2,flreview3,flreview4,flreview5,FloqTIReview,Asmar2024,Asmar2021}. Time-periodic electromagnetic fields hybridize the electronic and photon sectors, producing Floquet--Bloch bands absent in equilibrium, {while a choice of linearly (LPL) or circularly polarized light (CPL) can either preserve or selectively break time reversal symmetry to promote specific topological regimes}.

In this Letter, we show that commensurate type-II TMD heterobilayers { display non-trivial Floquet} topological {insulating phases} upon the incidence of near-infrared to visible {LPL or CPL}. Unlike previous studies on Floquet band engineering in TMDs\cite{Previous1,Previous2}, we investigate the near resonant regime, in which the one-photon valence bands cross the zero-photon conduction bands, mimicking static band inversion. Given the close spacing ($\sim 100\,{\rm meV}$) between same bands in opposite layers, as compared with the {relevant} band gaps ($\sim 1\,{\rm eV}$), this yields strong hybridization between four dominant Floquet bands, which we treat directly, and an additional four distant bands that we include perturbatively. 

We chart topological phase diagrams for the four {Floquet-Bloch} bands, computing their Chern numbers vs.\ the light frequency and field intensity, considering both parallel- (P) and anti-parallel (AP) commensurate heterobilayers at all {stable} interlayer registries. {Our calculations predict a narrow-gap QSHI phase for P stacking under LPL, and single-valley Chern insulator (CI) phases for AP stacking under CPL, with Chern numbers $\pm 2$ separated by band gaps of up to $20\,{\rm meV}$}. Our results establish TMD heterobilayers as an experimentally accessible platform for Floquet-engineered topological phases beyond the Landau-level paradigm.

\emph{Band models for TMD heterobilayers.} Consider a perfectly aligned, commensurate type-II TMD heterobilayer ${\rm MX}_2/{\rm M'X'}_2$, with in-plane atomic registry parametrized by the 2D vector $\rr_0$. In our convention, the highest valence bands at the $K$ and $K'=-K$  valleys belong to the bottom layer ${\rm MX}_2$, whereas the lowest conduction band belongs to the top layer ${\rm M'X'}_2$. A minimal model\cite{Tong2017} for the heterostructure spin-$s$ bands near the $\tau K$ point is\footnote{The basis ordering in Eq.\ \eqref{eq:4bands} is $\ket{c,\tau,s},\,\ket{v,\tau,s},\,\ket{c',\tau,s},\,\ket{v',\tau,s}$, where the primed (unprimed) bands correspond to the bottom (top) layer, and $c,\,v$ stand for conduction and valence, respectively. In our convention, the top valence band at valley $\tau=1$ has spin $\downarrow$.}
\begin{equation}\label{eq:4bands}
    H_{\tau,\kk,s}^{(0)} = \begin{pmatrix}
     \varepsilon_{c,\tau,s}& \gamma k_{-\tau} & t_{cc} & t_{cv} \\
    \gamma k_{\tau} & \varepsilon_{v,\tau,s} & t_{vc} & t_{vv} \\
    t_{cc} & t_{vc} & \varepsilon_{c',\zeta\tau,s} & \gamma' k_{-\zeta \tau} \\
    t_{cv} & t_{vv} & \gamma' k_{\zeta\tau} & \varepsilon_{v',\zeta\tau,s}
    \end{pmatrix},
\end{equation}
with $\tau=\pm1$, and $\varepsilon_g$ and $\varepsilon_g'$ the  $\hly$, and $\ely$ band gaps, respectively. Here, $k_\pm=k_x\pm i k_y$ are chiral BZ momenta, {$\zeta=1$ for P stacking (interlayer angle $\theta=0^\circ$), and $\zeta=-1$ for AP stacking ($\theta\,{\rm mod}120^\circ = 60^\circ$)}. The monolayer $\tau K$-point energies are
\begin{equation}\label{eq:energies}
\begin{split}
    \varepsilon_{c,\tau,s}=&-\tfrac{\Delta_0}{2} + \varepsilon_g + \tfrac{1+s\tau}{2}\Delta_c,\\
    \varepsilon_{v,\tau,s}=&-\tfrac{\Delta_0}{2}-\tfrac{1+s\tau}{2}\Delta_v,\\
    \varepsilon_{c',\tau\zeta,s}=&\tfrac{\Delta_0}{2}+\tfrac{1+s\tau\zeta}{2}\Delta_{c'},\\
    \varepsilon_{v',\tau\zeta,s}=&\tfrac{\Delta_0}{2} - \varepsilon_g' - \tfrac{1+s\tau\zeta}{2}\Delta_{v'},
\end{split}
\end{equation}
where $\Delta_c$ through $\Delta_{v'}$ are the spin-orbit splittings of the corresponding bands, and $\Delta_0$ determines the heterostructure gap. For simplicity, we have taken $\gamma$, $\gamma'$ and all interlayer tunnelling matrix elements $t_{\alpha\beta}$ as real. Importantly, all $t_{\alpha\beta}$ {depend} on both the stacking type $\zeta$ and the stacking vector $\rr_0$ through symmetry constraints\cite{Tong2017,landscapes,multifaceted} (Supplementary {Sec.}\ \ref{app:Lowdin}).

Since we are interested exclusively in the low-energy bands, we may project out the $\hly$ conduction and $\ely$ valence bands. We do so up to third order in perturbation theory using L\"owdin's partitioning\cite{lowdin1951,winkler} (Supplementary Sec.\ \ref{app:Lowdin}), resulting in the two-band model
\begin{equation}\label{eq:2bands}
    h_{\kk}=\begin{pmatrix}
    \varepsilon(k) + \nicefrac{\Delta(k)}{2} & t(\kk) \\
    t^*(\kk) & \varepsilon(k) - \nicefrac{\Delta(k)}{2}
    \end{pmatrix},
\end{equation}
where we have defined
\begin{equation}\label{eq:2bandparams}
\begin{split}
    \varepsilon(k) =& \varepsilon_0+\frac{\hbar^2k^2}{2m_-},\, \frac{\Delta(k)}{2} = \frac{\Delta_0 + \Delta_1}{2} + \frac{\hbar^2 k^2}{2m_+},\\
    t(\kk)=&t_{0} + \gamma_+k_+  + \gamma_-k_- + \gamma_1^2k^2 + \gamma_2^2 k_-^2.
\end{split}
\end{equation}
The coefficients in \eqref{eq:2bandparams} are given in terms of the four-band model \eqref{eq:4bands} parameters in Supplementary Sec.\ \ref{app:Lowdin}. In particular, $t_0$ through $\gamma_2$ directly depend on $t_{\alpha\beta}$, and as such they are stacking dependent. Upon band inversion, the topology of effective model \eqref{eq:2bands} depends on which term in $t(\kk)$ is dominant, yielding Chern numbers $0$, $\pm1$ or $\pm2$ when $t_0$, $k_\pm$ or $k_-^2$ dominate, respectively, achievable for different stackings, as show by Tong et al.\cite{Tong2017} (see Supplementary Table \ref{tab:parametersvsr0}).

The topological phases just described hinge on the feasibility of band inversion by static means. To date, the smallest band gap reported for a TMD hetero-bilayer is approximately\footnote{Reference \onlinecite{ju2024infrared} reports a photoluminescence peak in MoTe${}_2$/MoS${}_2$ at 0.8 eV, from the $K$-valley interlayer exciton. In Supplementary {Sec.}\ \ref{app:MoTe2}, we estimate an exciton binding energy of\cite{Danovich2018,rstarMoS2,rstarMoTe2,dinter1,dinter2,IXsPRB2020,Viner2021} $126\,{\rm meV}$, which suggests a heterostructure band gap of $0.926\,{\rm eV}$, in excellent agreement with the angle-resolved photoemission spectroscopy (ARPES) measurements of Ref.\ \onlinecite{2HMoTe2}.} $0.9\,{\rm eV}$ in MoTe${}_2$/MoS${}_2$ heterostructures\cite{2HMoTe2,ju2024infrared}. Tuning the heterostructure bandgap by almost a full electron-volt is close to the edge of current experimental capabilities using liquid ionic gating techniques\cite{ionicgating1_Morpurgo,ionicgating2_Bolotin,ionicgating3_Bolotin}. However, as we now discuss, a {dynamical} analogue of the required gap inversion is achievable through periodic energy pumping by an electromagnetic field of frequency $\Omega \sim \nicefrac{\Delta_0}{\hbar}$, in a much simpler experimental setting. 

\emph{Light-induced band inversion.} Let us now model the effects of an incident electromagnetic wave on the band-edge electrons. Considering a spatially uniform wave front, we perform the minimal substitution (Gaussian units) $\kk\rightarrow \kk
+\tfrac{e}{\hbar c}\AAA(t)$ in the four-band model \eqref{eq:4bands}, where $\AAA(t)$ is the time-dependent, Coulomb-gauge vector potential, $-e$ the electron charge, and $c$ the speed of light \emph{in vacuo}. We obtain
\begin{equation}\label{eq:4bandA}
    H_{\tau,\kk,s}(t)=H_{\tau,\kk,s}^{(0)}+\frac{e}{\hbar c}\begin{pmatrix}
        0 & \gamma A_{-\tau} & 0 & 0\\
        \gamma A_\tau & 0 & 0 & 0\\
        0 & 0 & 0 & \gamma' A_{-\zeta\tau}\\
        0 & 0 & \gamma' A_{\zeta\tau} & 0
    \end{pmatrix},
\end{equation}
with $A_\pm = A_x \pm i A_y$, reflecting the circular dichroism of TMDs\cite{Xiao2012}.

In the following, we consider the vector potential
{
\begin{equation}\label{eq:VecPot}
     \AAA(t) = \tfrac{cE_0}{\Omega}{\rm Re}\left\{ e^{i\Omega t}\left[ 
     \rho\left(\uvec{y}+i\uvec{x} \right) 
     + \lambda\left(-\uvec{y}+i\uvec{x} \right)e^{i\phi}\right]\right\},
\end{equation}
where $E_0$ is the electric-field amplitude and $\Omega$ is the driving frequency.
The two terms correspond to opposite circular polarizations and, with equal
weights ($\rho=\lambda=\nicefrac{1}{2}$), their superposition describes linearly polarized light (LPL).
The relative phase $\phi$ determines the polarization direction, with the angle
$\varphi$ measured from the $\uvec{x}$ axis given by
$\varphi=-\nicefrac{\phi}{2}$.
The corresponding right- and left-circularly polarized fields (RCPL and LCPL)
are obtained by retaining only the first ($\rho=\nicefrac{1}{2}$, $\lambda=0$) or second ($\rho=0$, $\lambda=\nicefrac{1}{2}$) circular component,
respectively.
}

\begin{figure}[t!]
    \centering
    \includegraphics[width=\columnwidth]{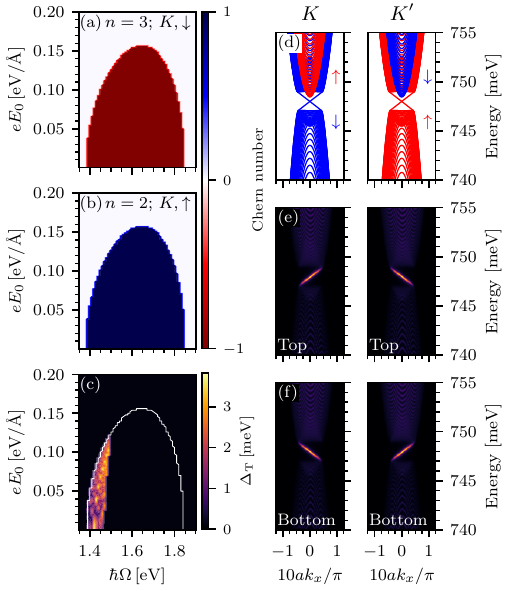}
    \caption{Topological phase diagram and chiral edge states of \PMX \,WSe${}_2$/MoSe${}_2$ driven with LPL. (a), (b) Chern numbers of $K$-point, spin-$\downarrow$ bulk bands $n=3$ and $n=2$, respectively, vs.\ field intensity and frequency. (c) Map of the topological band gap vs.\ field parameters, with the topological region delineated in white. (d) $K$ and $K'$ point band structures of a zig-zag nanoribbon of width $L=1500a$ for $\hbar\Omega=1.495\,{\rm eV}$ and $eE_0=60\,{\rm meV}\cdot\text{\AA}^{-1}$. (e) and (f) show the LDOS [$\log{(1+\rho)}$, bright indicates a high value] at the top and bottom edges of the ribbon for ($K,\,\downarrow$) bands on the left and ($K',\,\uparrow$) bands on the right.  The in-gap modes of panel (d) are spin-valley locked chiral edge states, a hallmark of QSHIs.}
    \label{fig:PhaseDiagramP}
\end{figure}

The Bloch Hamiltonian \eqref{eq:4bandA} defines a time-dependent Schr\"odinger equation (TDSE) $i\hbar \dot{u}_{\kk}(t) = H_{\kk}(t) u_{\kk}(t)$, which we solve using Floquet theory~\cite{Floquet1883,Shirley1965,Sambe1973}.
We write $u_{\kk}(t)=e^{-i\epsilon_{\kk} t/\hbar}\phi_{\kk}(t)$, where $\phi_{\kk}(t)$ is periodic in time with period $T=2\pi/\Omega$, and admits the Fourier expansion
$
    \phi_{\kk}(t)=\sum_{m\in\mathbb{Z}}e^{im\Omega t}\phi_{\kk;m}
$, 
turning the TDSE into the eigenvalue problem
\begin{equation}\label{eq:EigEq}
    \sum_{m'\in\mathbb{Z}}\left[\Hcal_{\kk;m,m'} + m\hbar\Omega\delta_{m,m'}\right]\phi_{\kk;m'} = \epsilon_{\kk} \phi_{\kk;m},
\end{equation}
with eigenvalues $\{\epsilon_{n,\kk}\}$ defining the {Floquet sidebands}. For monochromatic $H_{\kk}(t)$, the Floquet matrix elements $\Hcal_{\kk;m,m'} \equiv T^{-1}\int_0^{T}dt e^{-i(m-m')\Omega t}H_{\kk}(t)$ take the form
\begin{equation*}
    \Hcal_{\kk;m,m'}= H_{\kk}^{(0)}\delta_{m,m'}+ H_{\kk,+}\delta_{m,m'+1} + H_{\kk,-}\delta_{m,m'-1},
\end{equation*}
where $H_{\kk,+}=H_{\kk,-}^\dagger$. In our case, $H_{\kk,\pm}\equiv H_\pm$ are independent of $\kk$ and encode single-photon absorption/emission between neighboring Floquet sectors:
\begin{equation}
\begin{split}
    H_{-}&=-i\frac{eE_0}{2\hbar\Omega }\\
    &\times\begin{pmatrix}
        0 & \gamma f_{-\tau}(\phi) & 0 & 0\\
        \gamma f_{\tau}(\phi) & 0 & 0 & 0\\
        0 & 0 & 0 & \gamma' f_{-\zeta\tau}(\phi) \\
        0 & 0 & \gamma'f_{\zeta\tau}(\phi) & 0
    \end{pmatrix},
\end{split}
\end{equation}
where we have defined {$f_{-\tau}(\phi)=\rho(1+\tau)+\lambda(1-\tau)e^{-i\phi}$}.

Equation \eqref{eq:EigEq} is typically solved numerically by truncating the Floquet Hamiltonian to a finite set of harmonics. Here, we focus on frequencies $\hbar\Omega \gtrsim \Delta_0$ for which the equilibrium ($m=0$) bands cross their single-photon ($m=\pm1$) replicas, as in Fig.~\ref{fig:crossing}b. In this near-resonant regime, the quasienergy spectrum organizes into weakly coupled four-band sectors---e.g., two $m=0$ conduction bands hybridized with $m=1$ valence bands---while neighboring sectors are separated by energies of order $\Delta_0$. For weak light-matter coupling (photon-induced hybridizations $\ll \Delta_0$; see Supplementary {Sec.}\ \ref{app:field} and Refs.\ \cite{FloquetGraphene,Fluence} therein), higher-harmonic processes only weakly renormalize the sector band energies, so each sector can be treated independently to a good approximation~\cite{graphene-top-ins,gaps-floq-graphene,OnePh1,OnePh2,OnePh3}. We therefore restrict our analysis to the conduction-sector manifold near the Floquet-zone (FZ) edge ($\varepsilon\simeq\hbar\Omega/2$), formed by the two $m=0$ conduction bands and the $m=1$ replicas of the valence bands~\cite{graphene-top-ins,gaps-floq-graphene,OnePh1,OnePh2,OnePh3,Lauren2025}.

\emph{$P$ stacking}, $\uvec{x}$ \emph{linearly polarized light.} Near the FZ edge, the four-band conduction sector in the top half of Fig.\ \ref{fig:crossing}b is well described by Eq.\ \eqref{eq:EigEq}, taking only $m=0$ and $1$. The Floquet matrix is then of dimension $8$, including the resonant $m=0$ conduction- and $m=1$ valence bands, and the distant $m=0$ valence- and $m=1$ conduction bands. We derive effective models for the four resonant bands by projecting out the remote bands using L\"owdin's partitioning up to second order in perturbation theory. This gives the effective Hamiltonian for $P$-stacked bilayers
\begin{widetext}
\begin{equation}\label{eq:EffModelP}
    \tilde{\Hcal}^{{\rm P}}_{\tau,\kk,s}=\begin{pmatrix}
        \varepsilon_{c,\tau,s}+\tfrac{\gamma^2k^2}{\varepsilon_{c,\tau,s}-\varepsilon_{v,\tau,s}} & t_{12}+u_{12}^+k_\tau + u_{12}^-k_{-\tau}  & -\tfrac{ieE_0}{2\hbar\Omega}\gamma f_{-\tau}(\phi) & 0\\
        t_{12}^*+u_{12}^{+*}k_{-\tau} + u_{12}^{-*}k_\tau & \varepsilon_{c',\tau,s}+\tfrac{\gamma'{}^2k^2}{\varepsilon_{c',\tau,s}-\varepsilon_{v',\tau,s}} & 0 & -\tfrac{ieE_0 }{2\hbar\Omega}\gamma'f_{-\tau}(\phi) \\
        \tfrac{ieE_0}{2\hbar\Omega}\gamma f_{-\tau}^*(\phi) & 0 & \varepsilon_{v,\tau,s} + \hbar\Omega - \tfrac{\gamma^2k^2}{\varepsilon_{c,\tau,s}-\varepsilon_{v,\tau,s}} & t_{34} + u_{34}^+k_\tau +u_{34}^- k_{-\tau} \\
        0 & \tfrac{ieE_0 }{2\hbar\Omega}\gamma' f_{-\tau}^*(\phi) & t_{34}^*+u_{34}^{+*} k_{-\tau} + u_{34}^{-*}k_\tau & \varepsilon_{v',\tau,s} + \hbar\Omega - \tfrac{\gamma'{}^2k^2}{\varepsilon_{c',\tau,s}-\varepsilon_{v',\tau,s}}
    \end{pmatrix},
\end{equation}
\end{widetext}
written in the basis
\begin{equation*}
    \{\ket{c;m=0},\,\ket{c';m=0},\,\ket{v;m=1},\,\ket{v';m=1}\}.
\end{equation*}
The stacking-dependent model parameters of \eqref{eq:EffModelP} are reported in Supplementary Eq.\ \eqref{eq:EffModelP_params}.

The $P$ heterostructure exhibits circular dichroism, just like TMD monolayers. To see this, note that the Floquet coupling is always proportional to $f_{-\tau}(\phi)$, such that at valley $\tau=1$ ($\tau=-1$) only the RCPL (LCPL) component of the light field appears in the model. In the following we use LPL along the $\hat{\mathbf{x}}$ axis to introduce a Floquet coupling at both valleys, allowing for a true topological gap. For the sake of specificity, we take WSe${}_2$/MoSe${}_2$ as a representative case study, setting\cite{LatticeConstant1,LatticeConstant2,kormanyos2015k,Rivera_2015,Nagler_2017,Larentis,hX_Nature,Nguyen2019,PhysRevB.107.245407,Graham_2024} the lattice constant $a=3.289\,\text{\AA}$, $\gamma=2.6\,{\rm eV}\cdot \text{\AA}$, $\gamma'=2.2\,{\rm eV}\cdot \text{\AA}$, $\varepsilon_g=1.65\,{\rm eV}$, $\varepsilon_g'=1.58\,{\rm eV}$, $\Delta_0=1.38\,{\rm eV}$, $\Delta_c=12\,{\rm meV}$, $\Delta_{c'}=-37\,{\rm meV}$, $\Delta_v=500\,{\rm meV}$ and $\Delta_{v'}=220\,{\rm meV}$ (see Supplementary {Sec.}\ \ref{app:WSe2MoSe2parameters}).

The Chern number is the appropriate {topological invariant} for the irradiated system bands. Using the numerical method by Fukui et al.\cite{fukui2005chern}, we compute the Chern numbers of the resulting Floquet--Bloch quasienergy bands across a range of frequencies $\hbar\Omega \gtrsim \Delta_0$ and field intensities $eE_0$ of up to $200\,{\rm meV\cdot}\text{\AA}^{-1}$, and for all three stackings.\footnote{In all cases, the Chern numbers were computed in two approximations: first, using the Floquet-Bloch models as presented in Eqs.\ \eqref{eq:EffModelP} and \eqref{eq:EffModelAP} in a small region about the valley, and then in a regularized lattice scheme, where an entire ficticious BZ is considered. The two approaches agree for the parameter space explored. The presence of multiple parabolic bands with different curvatures can lead to spurious crossings. In our case, this happens for the $m=1$ valence bands, forcing us to introduce quartic terms in their energies that prevent the crossings, but which do not modify the Berry curvature in the region of interest.}. For \PMX \, stacking, the irradiated Floquet spectrum splits into topologically trivial and nontrivial phase space regions. In particular, bands $n=1$ and $n=4$ (as sorted by increasing energy) are topologically trivial across the entire parameter space explored, so we omit them in the following. By contrast, Figs.\ \ref{fig:PhaseDiagramP}a and \ref{fig:PhaseDiagramP}b show that the two central Floquet bands $n=2$ and 3 support a nontrivial {topological} phase diagram. This diagram consists of dome-shaped regions in the frequency-field strength plane with Chern numbers $\Ccal_2=1$ and $\Ccal_3=-1$ for the corresponding bands, characteristic of Floquet-induced band inversions in a continuum, periodically driven system. These domes terminate at a critical point $(\hbar\Omega,\,eE_0)\approx(1.64\,{\rm eV},\,160\,{\rm meV}/\text{\AA})$, above which both bands are topologically trivial.

Figure \ref{fig:PhaseDiagramP}c shows the topological band gap $\Delta_{\rm T}$ between the two middle bands, defined in terms of the ordinary gap $\Delta$ {between bands $n=2$ and 3} as
\begin{equation}\label{eq:topogap}
    \Delta_{\rm T} = \Delta\Theta(|\Ccal_1|+|\Ccal_2|+|\Ccal_3|+|\Ccal_4|),
\end{equation}
with $\Theta(x)$ the Heaviside theta function. Whenever $\Delta_{\rm T}$ is finite, bulk-edge correspondence for Floquet topological insulators predicts the emergence of chiral edge states traversing the quasienergy gap. To demonstrate this explicitly, we solved the effective model \eqref{eq:EffModelP} for a \PMX\, WSe${}_2$/MoSe${}_2$ nanoribbon of width $L\gg a$ with edges along the $\hat{\mathbf{x}}$ axis (zig-zag edges in an atomistic model, see Fig.\ \ref{fig:crossing}c), using the finite-differences approach of Ref.\ \onlinecite{Fedotov_2018}. Figure \ref{fig:PhaseDiagramP}d shows the corresponding quasienergy spectrum, revealing two counter-propagating modes connecting bands 2 and 3 across a true finite quasienergy gap of {about} $4\,{\rm meV}$ spanning both valleys. Given the time reversal symmetry inherited from the LPL vector potential, the heterostructure realizes a {periodically-driven} QSHI. Figures \ref{fig:PhaseDiagramP}e and \ref{fig:PhaseDiagramP}f show the local density of states (LDOS) computed at the top and bottom edges of the nanoribbon, respectively, showing that the in-gap modes of Fig.~\ref{fig:PhaseDiagramP}d are indeed chiral edge modes, as expected from bulk-edge correspondence. 

\begin{figure*}[t!]
    \centering
    \includegraphics[width=2\columnwidth]{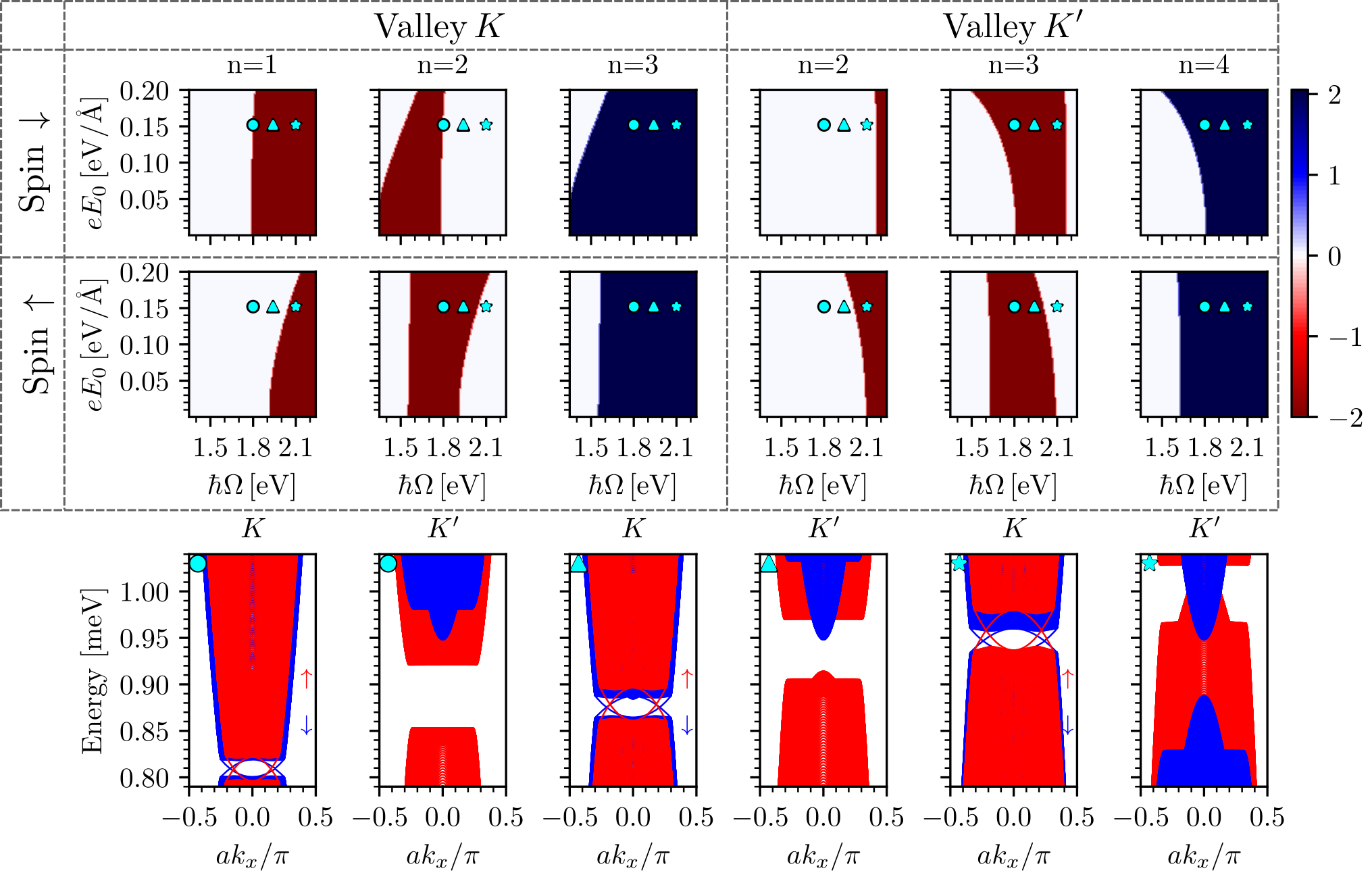}
    \caption{(Top) Topological phase diagram for 2H-WSe${}_2$/MoSe${}_2$ irradiated by RCPL, for both spins at both valleys. Bands with zero Chern number throughout the parameter space explored have been omitted. (Bottom) Band structures for a zig-zag nanoribbon ($L=1500a$) at the points of parameter space indicated with \textcolor{cyan}{$\bullet,\,\blacktriangle$} and \textcolor{cyan}{$\bigstar$} in the phase diagrams, corresponding to light-field parameters $\hbar\Omega=1.80,\,1.92$ and $2.1\,{\rm eV}$ for fixed $eE_0=156\,{\rm meV/\text{\AA}}$.}
    \label{fig:PhaseDiagramAP}
\end{figure*}

The nontrivial Floquet topology can be understood by projecting Eq.~\eqref{eq:EffModelP} onto the resonant $\{\ket{c';m=0},\,\ket{v;m=1}\}$ subspace governing the low-quasienergy structure near the FZ edge. These states are not directly coupled, but mix indirectly through light-field enabled virtual processes that visit the complementary subspace. For \PMX\, structures, the leading term going from $\ket{v;m=1}$ to $\ket{c;m=0}$ is proportional to $k_-$, leading to Chern numbers $\Ccal_2=-\Ccal_3=1$. Similarly, for \PXM\, stacking the dominant coupling scales as $k_+$, resulting in inverted Chern numbers $\Ccal_2=-\Ccal_3=-1$ (not shown). Both cases therefore realize {driven} QSHIs, with Chern numbers dynamically tunable at the FZ edge. It is noteworthy that \PMX\, and \PXM\, are stable stacking configurations for TMDs, and lead to the 3D bulk allotrope known as $3R$.

For completeness, we mention that in \PXX\, stacking all four bands are topologically trivial across the entire parameter space considered, so this case will not be discussed further.

\emph{2H stacking, circularly polarized light.} For AP bilayers, the minimal effective model that correctly reproduces the full-model band Chern numbers requires going to third order in perturbation theory.\footnote{The need to go to higher order stems from the opposite circular dichroisms of the $K$ and $K'$ valleys of the two layers, which coincide in AP structures. As a consequence, the momentum-independent terms in matrix elements $13$ and $24$ in Eq.\ \eqref{eq:EffModelAP} cannot be finite simultaneously. When one of these terms vanishes, the third-order contribution becomes the leading field-driven term, and as such it needs to be included in the effective model.} The resulting model is $\tilde{\Hcal}^{{\rm AP}}_{\tau,\kk,s}$, with matrix elements
\begin{equation}\label{eq:EffModelAP}
\begin{split}
    \left[\tilde{\Hcal}^{{\rm AP}}_{\tau,\kk,s} \right]_{11}=&\varepsilon_{c,\tau,s}+\frac{\gamma^2k^2}{\varepsilon_{c,\tau,s}-\varepsilon_{v,\tau,s}},\\
    \left[\tilde{\Hcal}^{{\rm AP}}_{\tau,\kk,s} \right]_{22}=&\varepsilon_{c',-\tau,s}+\frac{\gamma'{}^2k^2}{\varepsilon_{c',-\tau,s}-\varepsilon_{v',-\tau,s}},\\
    \left[\tilde{\Hcal}^{{\rm AP}}_{\tau,\kk,s} \right]_{33}=&\varepsilon_{v,\tau,s}+\hbar\Omega-\frac{\gamma^2k^2}{\varepsilon_{c,\tau,s}-\varepsilon_{v,\tau,s}},\\
    \left[\tilde{\Hcal}^{{\rm AP}}_{\tau,\kk,s} \right]_{44}=&\varepsilon_{v',-\tau,s}+\hbar\Omega-\frac{\gamma'{}^2k^2}{\varepsilon_{c',-\tau,s}-\varepsilon_{v',-\tau,s}},\\
    \left[\tilde{\Hcal}^{{\rm AP}}_{\tau,\kk,s} \right]_{12}=&s_{12}+v_{12}k_{-\tau} + w_{12}k^2+w_{12}^-k_{-\tau}^2,\\
    \left[\tilde{\Hcal}^{{\rm AP}}_{\tau,\kk,s} \right]_{13}=&-\frac{ieE_0}{2\hbar\Omega}\gamma f_{-\tau}(\phi)\\
    &+\frac{ieE_0}{(\hbar\Omega)^2}\frac{\gamma^3\left(\rho k_+ + \lambda e^{-i\phi}k_- \right)}{\varepsilon_{c,\tau,s}-\varepsilon_{v,\tau,s}}k_{-\tau},\\
    \left[\tilde{\Hcal}^{{\rm AP}}_{\tau,\kk,s} \right]_{14}=&v_{14}^-k_{-\tau} + v_{14}^+k_{\tau},\\
    \left[\tilde{\Hcal}^{{\rm AP}}_{\tau,\kk,s} \right]_{23}=&v_{23}^-k_{-\tau} + v_{23}^+ k_{\tau},\\
    \left[\tilde{\Hcal}^{{\rm AP}}_{\tau,\kk,s} \right]_{24}=&-\frac{ieE_0}{2\hbar\Omega}\gamma' f_{\tau}(\phi)\\
    &+\frac{ieE_0}{(\hbar\Omega)^2}\frac{\gamma'{}^3\left(\rho k_+ + \lambda e^{-i\phi}k_- \right)}{\varepsilon_{c',-\tau,s}-\varepsilon_{v',-\tau,s}}k_{\tau},\\
    \left[\tilde{\Hcal}^{{\rm AP}}_{\tau,\kk,s} \right]_{34}=&s_{34}+v_{34}k_{\tau}+w_{34}k^2 + w_{34}^+k_{\tau}^2,
\end{split}
\end{equation}
up to first order in $t_{cv}$ and $t_{vc}$. The model parameters are reported in Supplementary Eq.\ \eqref{eq:EffModelAP_params}. In the following we focus exclusively on the \APH\,case, since this is the only stable stacking for AP structures. 

We have numerically verified that LPL makes all bands topologically trivial for any angle $\varphi$ in 2H stacking. Instead, we present results for RCPL in Fig.\ \ref{fig:PhaseDiagramAP}. Corresponding results for LCPL are related by a time reversal operation, and are presented in Supplementary Fig.\ \ref{fig:2HLeft}. The top of Fig.\ \ref{fig:PhaseDiagramAP} shows the phase diagrams of all topologically nontrivial bands of both spins at either valley, exhibiting transitions from Chern number $0$ to $\pm2$. Inside a topological region for a given spin and valley, we expect chiral edge modes connecting the bulk bands of opposite Chern number. Figure \ref{fig:PhaseDiagramAP} shows that a topological gap exists between bands $n=2$ and $3$ only at the $K$ point, such that edge modes can only appear at this valley. This is confirmed numerically for a zig-zag nanoribbon at the bottom of Fig.\ \ref{fig:PhaseDiagramAP}, where two pairs of chiral edge bands per spin appear inside the $K$-valley gap at the phase space points marked $\bullet,\,\blacktriangle$ and $\bigstar$, corresponding to $\hbar\Omega=1.80,\,1.92$ and $2.1\,{\rm eV}$ for fixed $eE_0=156\,{\rm meV/\text{\AA}}$. By contrast, no in-gap states appear at the $K'$ point. LDOS calculations (Supplementary Fig.\ \ref{fig:2HLDOS}) show that the edge states of both spins inside the $K$-valley gap have the same chirality, leading to unidirectional, spin-neutral charge transport along the edge in the topological regime.

\begin{figure}[t!]
    \centering
    \includegraphics{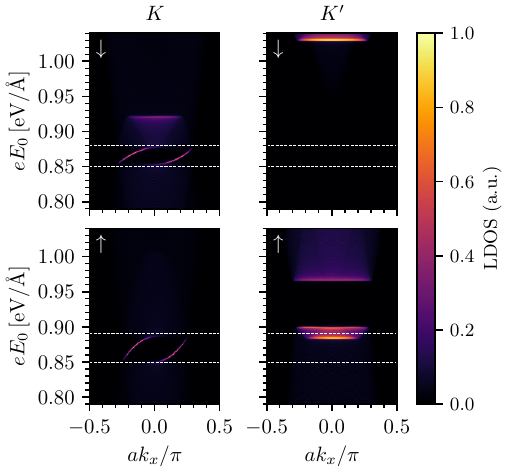}
    \caption{Spin- and valley-resolved LDOS (scaled as in Fig.\ \ref{fig:PhaseDiagramP}) at the top edge of nanoribbon considered in Fig.\ \ref{fig:PhaseDiagramAP}. Dashed lines frame the light-induced topological gap.}
    \label{fig:2HLDOSmain}
\end{figure}

As shown at the bottom of Fig.\ \ref{fig:PhaseDiagramAP}, the gaps at the two valleys never overlap to give a global topological gap. We have repeated the calculations for several $eE_0$ values across the topological region with the same results (Supplementary Fig.\ \ref{fig:2Hmorebands}). This indicates that pure edge transport shall remain obscured by the bulk states. However, these states can be specifically targeted by local probes, such as atomic resolution scanning tunneling microscopy (STM)\cite{ExpFor2H1,ExpFor2H2,ExpFor2H3,ExpFor2H4,ExpFor2H5,ExpFor2H6}. Figure \ref{fig:2HLDOSmain} for RCPL shows no LDOS support for $K'$-valley states at the ribbon edges within the energy window (of up to 20\,{\rm meV}, see Supplementary Fig.\ \ref{fig:2Hgap}) of the photoinduced gap, such that local, energy resolved measurements can specifically access the topologically protected edge modes.

\emph{Discussion.} The above results show that type-II TMD heterostructures may be turned into topological insulators upon periodic driving with IR to visible light of the appropriate polarization. In \PMX\,and\,\PXM\,structures, LPL preserves time reversal symmetry, inducing a Floquet QSHI phase with a topological gap of a few meV, detectable with TR-ARPES\cite{FloquetARPES1,FloquetARPES2,FloquetARPES3,FloquetARPES4,FloquetARPES5,FloquetARPES6}. In \APH\, structures both RCPL and LCPL drive a single-valley, high-Chern-number Chern insulator, featuring two pairs of chiral edge states per spin, accessible with local probe measurements\cite{ExpFor2H6}. 

Although TMD heterostructures are best known for exhibiting moir\'e patterns, single-stacking commensurate samples are also available experimentally. Weston et al.\cite{Reconstruction1} and Rosenberger et al.\cite{Reconstruction2} have measured strong reconstruction of marginally twisted heterobilayers, where mesoscopic ($\sim 100\,{\rm nm}$) \PMX\, and \APH\, domains are formed in chalcogen-matched structures. More recently, Baek et al.\cite{Commensurate1} have prepared fully commensurate samples, also with \PMX\, and \APH\, stackings, by annealing moir\'e samples encapsulated with hexagonal boron nitride or graphene.

Importantly, the emergence of higher Chern numbers $|\Ccal|=2$ does \emph{not} rely on the spatial modulation of light polarization, engineered polarization textures, or spatially structured irradiation to engineer higher Floquet topology~\cite{piskunow,Torres1}. Instead, it follows directly from near-resonant multiband hybridization in a spatially uniform drive\cite{UsajFloquet}. Taken together, these results establish the irradiated WSe${}_2$/MoSe${}_2$ heterobilayer as a genuinely driven topological system, whose Floquet--Bloch band topology is defined at the quasienergy level, and captured within a continuum description retaining the leading low-energy terms.

\acknowledgments{Work by Mahmoud M. Asmar was supported by the U.S. Department of Energy (DOE), Office of Science, Basic Energy Sciences (BES) under Award \# DE-SC0025703. E.S-S. and A.J.T.\ thank SECIHTI (Mexico) for financial support through its \emph{Becas Nacionales} Scholarship program. E.S-S., A.J.T.\ and D.A.R-T.\ acknowledge funding from PAPIIT-DGAPA-UNAM project IN114125. Finally, the authors acknowledge use of DeepSeek AI and Google Gemini as an advanced reference search tool. The authors independently retrieved, read, verified, and take full responsibility for the accuracy and integrity of all cited primary sources.}

\bibliographystyle{apsrev4-2}
\bibliography{biblio}

\renewcommand{\appendixname}{Supplementary Section}

\setcounter{figure}{0}
\renewcommand\thefigure{S\arabic{figure}}

\setcounter{table}{0}    
\renewcommand{\tablename}{Supplementary Table}

\newpage

\appendix

\begin{widetext}
\section{L\"owdin partitioning of the four-band model}\label{app:Lowdin}
In this section we obtain the model \eqref{eq:2bands} from \eqref{eq:4bands}, employing L\"owdin's partitioning\cite{lowdin1951} up to second order in perturbation theory. As we shall see, the two layers will are uncoupled at that order in the case of a P-stacked, MX' bilayer, and we are forced to go up to third order in perturbation theory for that case.

Briefly, L\"owdin's partitioning is designed to decouple two sectors $H_1$ and $H_2$ of a Hamiltonian $H=H_1+H_2+T$ up to arbitrary order in perturbation theory. Here, all matrix elements between blocks $H_1$ and $H_2$ are collected in $T$. Decoupling is achieved by means of a unitary transformation $H'=e^{-S}He^{S}$, with the generator $S$ some anti-Hermitian operator that can be expanded in powers of $T$ as $S=S^{(1)}+S^{(2)}+S^{(3)}+\ldots$ The form of $S^{(1)}$ to $S^{(\ell)}$ is chosen such that $H'=H_1'+H_2'+T'$, where $T'$ is at least of order $T^{\ell+1}$ and can be neglected, whereas the elements of $H_1'$ and $H_2'$ contain $H_1$ and $H_2$, respectively, plus corrections of order 2 to $\ell$ in $T$. If, say, subspace $H_1$ is of interest but not $H_2$, the resulting matrix $H_1'$ is an effective model for that subspace. A clear derivation of $S^{(\ell)}$ up to order $\ell=3$ is presented in Appendix B of Ref.\ \cite{winkler}. Here, we report only the corresponding expressions for the matrix elements $H_1'$, up to third order in $T$:
\begin{equation}\label{eq:lowdin}
\begin{split}
    &\braoket{m}{H_1'}{n}=\braoket{m}{H_1}{n} + \frac{1}{2}\sum_{l}^*\Big(\frac{\braoket{m}{T}{l}\braoket{l}{T}{n}}{\braoket{m}{H_1}{m}-\braoket{l}{H_2}{l}} + \frac{\braoket{m}{T}{l}\braoket{l}{T}{n}}{\braoket{n}{H_1}{n}-\braoket{l}{H_2}{l}} \Big)\\
    &-\frac{1}{2}\sum_{l}^*\sum_{p}^{\star}\left(\frac{\braoket{m}{T}{l}\braoket{l}{T}{p}\braoket{p}{T}{n}}{\left(\braoket{n}{H_1}{n} - \braoket{l}{H_2}{l}  \right)\left(\braoket{p}{H_1}{p} - \braoket{l}{H_2}{l}  \right)} + \frac{\braoket{m}{T}{p}\braoket{p}{T}{l}\braoket{l}{T}{n}}{\left(\braoket{m}{H_1}{m} - \braoket{l}{H_2}{l}  \right)\left(\braoket{p}{H_1}{p} - \braoket{l}{H_2}{l}  \right)} \right)\\
    &+\frac{1}{2}\sum_{l,l'}^*\left(\frac{\braoket{m}{T}{l}\braoket{l}{T}{l'}\braoket{l'}{T}{n}}{\left( \braoket{m}{H_1}{m}-\braoket{l}{H_2}{l}\right)\left( \braoket{m}{H_1}{m}-\braoket{l'}{H_2}{l'}\right)}+ \frac{\braoket{m}{T}{l}\braoket{l}{T}{l'}\braoket{l'}{T}{n}}{\left( \braoket{n}{H_1}{n}-\braoket{l}{H_2}{l}\right)\left( \braoket{n}{H_1}{n}-\braoket{l'}{H_2}{l'}\right)} \right).
\end{split}
\end{equation}
Naturally, the indices $m,\,n$ belong to block $H_1$. The $*$ in the sums indicates that $l$ and $l'$ run exclusively over block $H_2$, whereas $\star$ indicates that $p$ runs over block $H_1$. 

Applying Eq.\ \eqref{eq:lowdin} to (1)  and choosing $H_1$ as the $\{\ket{\ely;c},\,\ket{\hly;v}\}$ subspace, where $c$ and $v$ represent conduction and valence, we obtain the following effective models:
{\small
\begin{align}\label{eq:models}
h_{\kk}^{P,XX'}=&\left[\left(\frac{|t_{vv}|^2}{2(\varepsilon_g'-\Delta_0)} - \frac{|t_{cc}|^2}{2(\varepsilon_g-\Delta_0)}\right) +\frac{1}{2}\left(\frac{|\gamma'|^2}{\varepsilon_g'} - \frac{|\gamma|^2}{\varepsilon_g}\right)q^2 \right]\sigma_0 \nonumber\\
&+\frac{1}{2}\left[\Delta_0-\left(\frac{|t_{vv}|^2}{(\varepsilon_g'-\Delta_0)} + \frac{|t_{cc}|^2}{(\varepsilon_g-\Delta_0)}\right) +  \left(\frac{|\gamma'|^2}{\varepsilon_g'} + \frac{|\gamma|^2}{\varepsilon_g}\right)q^2 \right]\sigma_z\nonumber\\
&+\mathrm{Re}\left[\left( \frac{(2\varepsilon_g'-\Delta_0)\gamma't_{vv}^*}{2\varepsilon_g'(\varepsilon_g'-\Delta_0)}-\frac{(2\varepsilon_g-\Delta_0)\gamma t_{cc}^*}{2\varepsilon_g(\varepsilon_g-\Delta_0)} \right)q_- \right]\sigma_x\nonumber\\
&-\mathrm{Im}\left[\left( \frac{(2\varepsilon_g'-\Delta_0)\gamma't_{vv}^*}{2\varepsilon_g'(\varepsilon_g'-\Delta_0)}-\frac{(2\varepsilon_g-\Delta_0)\gamma t_{cc}^*}{2\varepsilon_g(\varepsilon_g-\Delta_0)} \right)q_- \right]\sigma_y,\nonumber\\
h_{\kk}^{P,MX'}=&\left[\frac{1}{2}\left(\frac{|\gamma'|^2}{\varepsilon_g'} - \frac{|\gamma|^2}{\varepsilon_g}\right)q^2 \right]\sigma_0+\frac{1}{2}\left[\Delta_0+\left(\frac{|\gamma'|^2}{\varepsilon_g'} + \frac{|\gamma|^2}{\varepsilon_g}\right)q^2\right]\sigma_z\nonumber\\
&+ \mathrm{Re}\left[ -\frac{\gamma\gamma' t_{cv^*}}{2}\left(\frac{1}{\varepsilon_g(\varepsilon_g'-\Delta_0)} + \frac{1}{\varepsilon_g'(\varepsilon_g-\Delta_0)} \right)q_-^2 \right]\sigma_x\nonumber\\
&- \mathrm{Im}\left[ -\frac{\gamma\gamma' t_{cv^*}}{2}\left(\frac{1}{\varepsilon_g(\varepsilon_g'-\Delta_0)} + \frac{1}{\varepsilon_g'(\varepsilon_g-\Delta_0)} \right)q_-^2 \right]\sigma_y,\nonumber\\
h_{\kk}^{P,XM'}=&\left[ \frac{1}{2}\left(\frac{|\gamma'|^2}{\varepsilon_g'} - \frac{|\gamma|^2}{\varepsilon_g}\right)q^2 \right]\sigma_0+\frac{1}{2}\left[\Delta_0+\left(\frac{|\gamma'|^2}{\varepsilon_g'} + \frac{|\gamma|^2}{\varepsilon_g}\right)q^2\right]\sigma_z\nonumber\\
&+ \mathrm{Re}\left[t_{vc}^* - \frac{t_{vc}^*}{2}\left(\frac{|\gamma|^2}{\varepsilon_g(\varepsilon_g-\Delta_0)}  + \frac{|\gamma'|^2}{\varepsilon_g'(\varepsilon_g'-\Delta_0)} \right)q^2 \right]\sigma_x\nonumber\\
&-  \mathrm{Im}\left[t_{vc}^* - \frac{t_{vc}^*}{2}\left(\frac{|\gamma|^2}{\varepsilon_g(\varepsilon_g-\Delta_0)}  + \frac{|\gamma'|^2}{\varepsilon_g'(\varepsilon_g'-\Delta_0)} \right)q^2 \right]\sigma_y,\nonumber\\
h_{\kk}^{AP, XX'}=&\left[ \frac{1}{2}\left(\frac{|\gamma'|^2}{\varepsilon_g'} - \frac{|\gamma|^2}{\varepsilon_g}\right)q^2 \right]\sigma_0+\frac{1}{2}\left[\Delta_0+\left(\frac{|\gamma'|^2}{\varepsilon_g'} + \frac{|\gamma|^2}{\varepsilon_g}\right)q^2\right]\sigma_z\nonumber\\
&+ \mathrm{Re}\left[t_{vc}^* - \left( \frac{t_{vc}^*}{2}\left(\frac{|\gamma|^2}{\varepsilon_g(\varepsilon_g-\Delta_0)} + \frac{|\gamma'|^2}{\varepsilon_g'(\varepsilon_g'-\Delta_0)} \right) + \frac{t_{cv}^*}{2}\gamma\gamma'{}^*\frac{2\varepsilon_g\varepsilon_g'-(\varepsilon_g+\varepsilon_g')\Delta_0}{\varepsilon_g\varepsilon_g'(\varepsilon_g-\Delta_0)(\varepsilon_g'-\Delta_0)} \right)q^2 \right]\sigma_x\nonumber\\
&-  \mathrm{Im}\left[t_{vc}^* - \left( \frac{t_{vc}^*}{2}\left(\frac{|\gamma|^2}{\varepsilon_g(\varepsilon_g-\Delta_0)} + \frac{|\gamma'|^2}{\varepsilon_g'(\varepsilon_g'-\Delta_0)} \right) + \frac{t_{cv}^*}{2}\gamma\gamma'{}^*\frac{2\varepsilon_g\varepsilon_g'-(\varepsilon_g+\varepsilon_g')\Delta_0}{\varepsilon_g\varepsilon_g'(\varepsilon_g-\Delta_0)(\varepsilon_g'-\Delta_0)} \right)q^2 \right]\sigma_y,\nonumber\\
h_{\kk}^{AP,MM'}=&\left[-\frac{|t_{cc}|^2}{2(\varepsilon_g-\Delta_0)} + \frac{1}{2}\left(\frac{|\gamma'|^2}{\varepsilon_g'} - \frac{|\gamma|^2}{\varepsilon_g}\right)q^2 \right]\sigma_0+\frac{1}{2}\left[\Delta_0-\frac{|t_{cc}|^2}{(\varepsilon_g-\Delta_0)}+\left(\frac{|\gamma'|^2}{\varepsilon_g'} + \frac{|\gamma|^2}{\varepsilon_g}\right)q^2\right]\sigma_z\nonumber\\
&+\mathrm{Re}\left[-\frac{(2\varepsilon_g-\Delta_0)\gamma t_{cc}^*}{2\varepsilon_g(\varepsilon_g-\Delta_0)}q_- \right]\sigma_x -\mathrm{Im}\left[-\frac{(2\varepsilon_g-\Delta_0)\gamma t_{cc}^*}{2\varepsilon_g(\varepsilon_g-\Delta_0)}q_- \right]\sigma_y,\nonumber\\
h_{\kk}^{AP,2H}=&\left[\frac{|t_{vv}|^2}{2(\varepsilon_g'-\Delta_0)} + \frac{1}{2}\left(\frac{|\gamma'|^2}{\varepsilon_g'} - \frac{|\gamma|^2}{\varepsilon_g} \right)q^2\right]\sigma_0+\frac{1}{2}\left[\Delta_0 - \frac{|t_{vv}|^2}{(\varepsilon_g'-\Delta_0)} + \left(\frac{|\gamma'|^2}{\varepsilon_g'} + \frac{|\gamma|^2}{\varepsilon_g} \right)q^2 \right]\sigma_z\nonumber\\
&+\mathrm{Re}\left[\frac{(2\varepsilon_g'-\Delta_0)\gamma'{}^* t_{vv}}{2\varepsilon_g'(\varepsilon_g'-\Delta_0)}q_+ \right]\sigma_x -\mathrm{Im}\left[\frac{(2\varepsilon_g'-\Delta_0)\gamma'{}^* t_{vv}}{2\varepsilon_g'(\varepsilon_g'-\Delta_0)}q_+ \right]\sigma_y,
\end{align}}
where we have introduced the Pauli matrices, acting on the electron-layer conduction, hole-layer valence band subspace. These models have already been simplified by considering the symmetry constraints imposed on the interlayer tunnelling energies $t_{cc},\,t_{vv},\,t_{cv}$ and $t_{vc}$. Briefly, $t_{cv}=t_{vc}=0$ for (P, XX'); $t_{cc}=t_{vc}=t_{vv}=0$ for (P, MX'); $t_{cc}=t_{cv}=t_{vv}=0$ for (P, XM'); $t_{cc}=t_{vv}=0$ for (AP, XX'); $t_{vv}=t_{cv}=t_{vc}=0$ for (AP, MM'); and $t_{cc}=t_{vc}=t_{cv}=0$ for (AP, 2H). These constraints are justified by Tong et al. in Ref.\ \onlinecite{Tong2017}, based on tight-binding arguments. An alternative proof is presented in Refs.\ \onlinecite{landscapes,multifaceted}, based on a L\"owdin-ortho-normalized hybrid $\kk\cdot\pp$-tight-binding model. The models \eqref{eq:models} where first reported in Ref.\ \onlinecite{Tong2017}, albeit simplified by the assumption $\Delta_0\ll\varepsilon_g,\,\varepsilon_g'$.

The model parameters of Eq.\ \eqref{eq:models} are summarized in Table \ref{tab:parametersvsr0}, where further details on the specific stacking configurations are also listed.  Table \ref{tab:parametersvsr0} also lists the valence-band Chern number $\Ccal_v$ obtained upon band inversion ($\Delta < 0$) using the effective 2-band model \eqref{eq:2bands} for each configuration, with $\Ccal_c = -\Ccal_v$ for the conduction band. We have confirmed that the same Chern numbers are obtained from the full $4$-band model \eqref{eq:4bands}.

\begin{table*}[h!]
\caption{Leading-order parameters of effective model \eqref{eq:2bands} at high-symmetry stacking configurations in parallel- (P) and anti parallel (AP) stacked bilayers. In each case, $\rr_0$ is written in terms of the lattice vectors $\aaa_1=a(1,\,0)$ and $\aaa_2=a(\nicefrac{1}{2},\,\nicefrac{\sqrt{3}}{2})$, with $a$ the lattice constant. When sub-leading-order parameters are finite, they are indicated by ``$-$''. $\Ccal_v$ is the valence band Chern number, and $\Ccal_c = -\Ccal_v$ in all cases.}
\begin{center}
\begin{tabular}{l    c | c c c  c c  r}
\hline\hline
P stacking  & Registry vector $\rr_0$  & $t_0$ & $\gamma_+$ & $\gamma_-$ & $\gamma_1^2$ & $\gamma_2^2$&$\Ccal_v$\\
\hline
${\rm XX'}$ & $\boldsymbol{0}$     &       0&       0&$\tfrac{\gamma't_{vv}^*(2\varepsilon_g'-\Delta_0)}{2\varepsilon_g'(\varepsilon_g'-\Delta_0)}-\tfrac{\gamma t_{cc}^*(2\varepsilon_g-\Delta_0)}{2\varepsilon_g(\varepsilon_g-\Delta_0)}$&0&0&$1$ \\
${\rm MX'}$ & $\nicefrac{(\aaa_1-2\aaa_2)}{3}$   & 0&0&0&0&$\tfrac{-\gamma\gamma' t_{cv^*}}{2\varepsilon_g(\varepsilon_g'-\Delta_0)} + \tfrac{-\gamma\gamma' t_{cv^*}}{2\varepsilon_g'(\varepsilon_g-\Delta_0)}$&$2$ \\
${\rm XM'}$ & $\nicefrac{(2\aaa_2-\aaa_1)}{3}$   & $t_{vc}^*$&0&0&--&0&0 \\
\hline\hline
AP stacking & Registry vector $\rr_0$   & $t_0$ & $\gamma_+$ & $\gamma_-$ & $\gamma_1^2$ & $\gamma_2^2$&$\Ccal_v$\\
\hline
${\rm XX'}$ & $\nicefrac{(2\aaa_2-\aaa_1)}{3}$ & $t_{vc}^*$&0&0&--&0&$0$ \\
${\rm MM'}$ & $\nicefrac{(\aaa_1-2\aaa_2)}{3}$& 0 &     0&$-\tfrac{\gamma t_{cc}^*(2\varepsilon_g-\Delta_0)}{2\varepsilon_g(\varepsilon_g-\Delta_0)}$&0&0&$1$ \\
${\rm 2H}$ & $\boldsymbol{0}$ &0&  $\tfrac{\gamma'{}^*t_{vv}(2\varepsilon_g'-\Delta_0)}{2\varepsilon_g'(\varepsilon_g'-\Delta_0)}$&0&0&0&$-1$ \\
\hline\hline
\multicolumn{2}{l}{All cases} & \multicolumn{3}{c}{$m_-=\hbar^2\left(\tfrac{|\gamma'|^2}{\varepsilon_g'} - \tfrac{|\gamma|^2}{\varepsilon_g}  \right)^{-1}$ } & \multicolumn{3}{c}{$m_+=\hbar^2\left(\tfrac{|\gamma'|^2}{\varepsilon_g'} + \tfrac{|\gamma|^2}{\varepsilon_g}  \right)^{-1}$ }
\\
\hline\hline
\end{tabular}\\
\end{center}
\label{tab:parametersvsr0}
\end{table*}

\section{Estimate of the ${\rm MoTe}_2/{\rm MoS}_2$ band gap}\label{app:MoTe2}
Reference \onlinecite{ju2024infrared} reports an interlayer exciton energy of $0.8\,{\rm eV}$ in the heterobilayer MoTe${}_2$/MoS${}_2$. This energy relates to the single-particle heterostructure band gap $\varepsilon_g$ as
\begin{equation*}
    \varepsilon_{\rm IX}=\varepsilon_g - \varepsilon_b,
\end{equation*}
where $\varepsilon_b$ is the exciton binding energy. We have computed $\varepsilon_b$ by solving the two-dimensional Wannier-Mott equation for this system, considering a MoS${}_2$ electron with effective mass $m_{\rm e}=0.61\,{\rm m_0}$ and a MoTe${}_2$ hole with effective mass $m_{\rm h}=1.17\,m_0$, $m_0$ being the bare electron mass, with a permanent vertical separation $d$. Following Ref.\ \onlinecite{Danovich2018}, we take an electron-hole interaction potential of the form
\begin{equation*}
\begin{split}
    V_{\rm e-h}(r)=&4\pi\int dq\,q\,V_{\rm e-h}(q),\\
    V_{\rm e-h}(q)=&\frac{2\pi e^{-q d}}{\epsilon q \left[(1+r_{\rm S}q)(1+r_{\rm T}q)-r_{\rm S}r_{\rm T}q^2 e^{-2q d}\right]},
\end{split}
\end{equation*}
where $r_{\rm X}=2\pi\kappa_{\rm X}/\tilde\epsilon$ is known as the screening length of the MoX${}_2$ layer, defined in terms of its in-plane dielectric polarizability $\kappa_{\rm X}$, and $\epsilon$ is the dielectric constant of the environment. To simulate the SiO${}_2$ substrate used in the samples of Ref.\ \onlinecite{ju2024infrared}, we take the average dielectric constant $\epsilon=(1+\epsilon_{\rm SiO_2})/2=2.4$ of the substrate below and vacuum above. From Refs.\ \onlinecite{rstarMoS2} and \onlinecite{rstarMoTe2} we get $r_{\rm S} \epsilon=38.62\,\text{\AA}$ and $r_{\rm T} \epsilon=73.61\,\text{\AA}$. Finally, we took $d=6.574\,\text{\AA}$, the average of the bulk interlayer distances of MoS${}_2$ and MoTe${}_2$.\cite{dinter1,dinter2} 

With the pontential and parameters discused, we solved the Wannier-Mott equation numerically using the method discussed in Refs.\ \onlinecite{IXsPRB2020} and \onlinecite{Viner2021}. We find a binding energy $\varepsilon_b=126\,{\rm meV}$ for the $1s$ exciton, leading to an estimated heterostructure band gap $\varepsilon_g=0.926\,{\rm eV}$.

\section{Electric fields within current experimental capabilities}\label{app:field}
To set an upper limit on the field strengths considered in our calculations, we look at the relationship between field strength and laser fluence reported in the recent experimental literature. Reference \cite{FloquetGraphene} reports first-order Floquet gaps of approximately $10$ and $ 60\,{\rm meV}$ in graphene pumped with  a $500\,{\rm fs}$ pulsed laser, at a frequency of $\hbar\Omega = 200\,{\rm meV}$, and for laser fluences of $0.03$ and $0.23\,{\rm mJ}\cdot {\rm cm}^{-2}$, respectively. The relation between these gaps $\delta$ and the electric field strengths $eE_0$ attained at the sample can be estimated as
\begin{equation*}
    \delta = \left|\frac{eE_0}{\hbar\Omega}\gamma_{\rm G}\right|,
\end{equation*}
with $|\gamma_{\rm G}|=\hbar v_F = 6.6\,{\rm eV}\cdot \text{\AA}$ is the light-matter coupling constant in graphene. From this we estimate field strengths of $0.27$ and $1.8\,{\rm meV}\cdot\text{\AA}^{-1}$, corresponding to the above cited fluences. Assuming a linear relation between field strength $eE_0$ and fluence $f$, we obtain
\begin{equation}\label{eq:eE0_vs_f}
    eE_0 \approx 7.7\,\frac{{\rm meV}\cdot\text{\AA}^{-1}}{{\rm mJ}\cdot {\rm cm}^{-1}} f,
\end{equation}
with $f$ in units of ${\rm mJ}\cdot {\rm cm}^{-1}$.

Reference \cite{Fluence} reports the use of $160\,{\rm fs}$ pulsed lasers with an $800\,{\rm nm}$ wavelength, or $\hbar\Omega = 1.55\,{\rm eV}$---right within the range of frequencies required to explore the topological phases described in the main text---on MoS${}_2$ sitting on several typical substrates. The authors report no ablation for the sample, using fluences ranging from 20 to $400\,{\rm mJ}\cdot{\rm cm}^{-2}$, setting a clear experimental upper limit. Equation \eqref{eq:eE0_vs_f} suggests that these fluences give electric fields ranging from $154\,{\rm meV}\cdot\text{\AA}^{-1}$ to $3.1\,{\rm eV}\cdot\text{\AA}^{-1}$.

In the main text we consider a conservative upper limit of $200\,{\rm meV}\cdot\text{\AA}^{-1}$, corresponding to a laser fluence of about $30\,{\rm mJ}\cdot{\rm cm}^{-2}$.

\section{Effective Bloch Hamiltonians at $\hbar\Omega\gtrsim \Delta_0$}\label{app:4bandParams}
Fully expanded, the matrix elements of Eq.\ \eqref{eq:EffModelP} are
\begin{equation}\label{eq:EffModelPExpanded}
    \tilde{\Hcal}_{{\rm P},\kk}=\begin{pmatrix}
        \varepsilon_{c,\tau,s}+\tfrac{\gamma^2k^2}{\varepsilon_{c,\tau,s}-\varepsilon_{v,\tau,s}} & t_{12}+u_{12}^+k_\tau + u_{12}^-k_{-\tau}  & -\tfrac{ieE_0}{2\hbar\Omega}\gamma f_{-\tau}(\phi) & 0\\
        t_{12}^*+u_{12}^{+*}k_{-\tau} + u_{12}^{-*}k_\tau & \varepsilon_{c',\tau,s}+\tfrac{\gamma'{}^2k^2}{\varepsilon_{c',\tau,s}-\varepsilon_{v',\tau,s}} & 0 & -\tfrac{ieE_0 }{2\hbar\Omega}\gamma'f_{-\tau} \\
        \tfrac{ieE_0}{2\hbar\Omega}\gamma f_{-\tau}^*(\phi) & 0 & \varepsilon_{v,\tau,s} + \hbar\Omega - \tfrac{\gamma^2k^2}{\varepsilon_{c,\tau,s}-\varepsilon_{v,\tau,s}} & t_{34} + u_{34}^+k_\tau +u_{34}^- k_{-\tau} \\
        0 & \tfrac{ieE_0 }{2\hbar\Omega}\gamma' f_{-\tau}^*(\phi) & t_{34}^*+u_{34}^{+*} k_{-\tau} + u_{34}^{-*}k_\tau & \varepsilon_{v',\tau,s} + \hbar\Omega - \tfrac{\gamma'{}^2k^2}{\varepsilon_{c',\tau,s}-\varepsilon_{v',\tau,s}}
    \end{pmatrix}.
\end{equation}

We report the parameters of models \eqref{eq:EffModelP} and \eqref{eq:EffModelAP}, as obtained from the second- and third-order L\"owding partitionings of the full Floquet-Bloch eigenvalue problem \eqref{eq:EigEq}. The P-stacking model \eqref{eq:EffModelP} contains the six parameters
\begin{equation}\label{eq:EffModelP_params}
\begin{split}
    &t_{12}=\left\{\begin{array}{rl}
         t_{cc},& \mathrm{XX'}  \\
         0,& \mathrm{MX'}\\
         0,& \mathrm{XM'}
    \end{array}\right.,\quad u_{12}^+=\left\{\begin{array}{rl}
         0,& \mathrm{XX'}  \\
         \tfrac{\gamma't_{cv}}{2}\left(\tfrac{1}{\varepsilon_{c,\tau,s}-\varepsilon_{v,\tau,s}}+\tfrac{1}{\varepsilon_{c',\tau,s}-\varepsilon_{v',\tau,s}} \right),& \mathrm{MX'}\\
         0,& \mathrm{XM'}
    \end{array}\right.,\\
    &u_{12}^-=\left\{\begin{array}{rl}
         0,& \mathrm{XX'}  \\
         0,& \mathrm{MX'}\\
         \tfrac{\gamma t_{vc}}{2}\left(\tfrac{1}{\varepsilon_{c,\tau,s}-\varepsilon_{v,\tau,s}}+\tfrac{1}{\varepsilon_{c',\tau,s}-\varepsilon_{v',\tau,s}} \right),& \mathrm{XM'}
    \end{array}\right.,\quad t_{34}=\left\{\begin{array}{rl}
         t_{vv},& \mathrm{XX'}  \\
         0,& \mathrm{MX'}\\
         0,& \mathrm{XM'}
    \end{array}\right.,\\
    &u_{34}^+=\left\{\begin{array}{rl}
         0,& \mathrm{XX'}  \\
         -\tfrac{\gamma t_{cv}}{2}\left(\tfrac{1}{\varepsilon_{c,\tau,s}-\varepsilon_{v,\tau,s}} + \tfrac{1}{\varepsilon_{c',\tau,s}-\varepsilon_{v',\tau,s}} \right),& \mathrm{MX'}\\
         0,& \mathrm{XM'}
    \end{array}\right.,\,u_{34}^-=\left\{\begin{array}{rl}
         0,& \mathrm{XX'}  \\
         0,& \mathrm{MX'}\\
         -\tfrac{\gamma' t_{vc}}{2}\left(\tfrac{1}{\varepsilon_{c,\tau,s}-\varepsilon_{v,\tau,s}} + \tfrac{1}{\varepsilon_{c',\tau,s}-\varepsilon_{v,\tau,s}} \right),& \mathrm{XM'}
    \end{array}\right..
\end{split}
\end{equation}
Conversely, the AP-stacking model \eqref{eq:EffModelP} contains the twelve parameters
\begin{align}\label{eq:EffModelAP_params}
    &s_{12}=\left\{\begin{array}{rl}
         0,& \mathrm{XX'}  \\
         t_{cc},& \mathrm{MM'}\\
         0,& \mathrm{2H}
    \end{array} \right.,\,v_{12}=\left\{\begin{array}{rl}
         \tfrac{\gamma t_{vc}}{2}\left(\tfrac{1}{\varepsilon_{c,\tau,s}-\varepsilon_{v,\tau,s}}+\tfrac{1}{\varepsilon_{c',-\tau,s}-\varepsilon_{v,\tau,s}} \right),& \mathrm{XX'}\\+\tfrac{\gamma' t_{cv}}{2}\left(\tfrac{1}{\varepsilon_{c,\tau,s}-\varepsilon_{v',-\tau,s}}+\tfrac{1}{\varepsilon_{c',-\tau,s}-\varepsilon_{v',-\tau,s}} \right)  \\
         0,& \mathrm{MM'}\\
         0,& \mathrm{2H}
    \end{array} \right.,\nonumber\\ 
    &w_{12}=\left\{\begin{array}{rl}
         0,& \mathrm{XX'}  \\
         -\tfrac{t_{cc}}{2}\left[\tfrac{\gamma^2}{\left(\varepsilon_{c,\tau,s}-\varepsilon_{v,\tau,s} \right)\left(\varepsilon_{c',-\tau,s}-\varepsilon_{v,\tau,s} \right)} + \tfrac{\gamma'{}^2}{\left(\varepsilon_{c,\tau,s}-\varepsilon_{v',-\tau,s} \right)\left(\varepsilon_{c',-\tau,s}-\varepsilon_{v',-\tau,s} \right)} \right],& \mathrm{MM'}\\
         0,& \mathrm{2H}
    \end{array} \right.,\nonumber\\
    &w_{12}^-=\left\{\begin{array}{rl}
         0,& \mathrm{XX'}  \\
         0,& \mathrm{MM'}\\
         \tfrac{\gamma\gamma' t_{vv}}{2}\left[\tfrac{1}{\left(\varepsilon_{c,\tau,s}-\varepsilon_{v,\tau,s}\right)\left(\varepsilon_{c,\tau,s}-\varepsilon_{v',-\tau,s} \right)} + \tfrac{1}{\left(\varepsilon_{c',-\tau,s}-\varepsilon_{v,\tau,s} \right)\left(\varepsilon_{c',-\tau,s}-\varepsilon_{v',-\tau,s} \right)} \right],& \mathrm{2H}
    \end{array} \right.,\nonumber\\
    &v_{14}^-=\left\{\begin{array}{rl}
         -\tfrac{ieE_0 f_{\tau}(\phi)}{(2\hbar\Omega)^2}\Bigg[\tfrac{\gamma\gamma' \hbar\Omega t_{vc}}{\left(\varepsilon_{v,\tau,s}-\varepsilon_{v',-\tau,s}-\hbar\Omega \right)\left(\varepsilon_{c',-\tau,s}-\varepsilon_{v,\tau,s} \right)}-\tfrac{\gamma'{}^2 t_{cv}}{\varepsilon_{c',-\tau,s}-\varepsilon_{v',-\tau,s}} & \mathrm{XX'} \\ +\tfrac{\gamma^2 \hbar\Omega t_{cv}}{\left(\varepsilon_{c,\tau,s}-\varepsilon_{v',-\tau,s}\right)\left(\varepsilon_{v,\tau,s}-\varepsilon_{v',-\tau,s}-\hbar\Omega \right)}-\tfrac{\gamma^2 t_{cv}}{\varepsilon_{c,\tau,s}-\varepsilon_{v,\tau,s}} \Bigg],  \\
         0,& \mathrm{MM'}\\
         0,& \mathrm{2H}
    \end{array} \right.,\nonumber\\
    &v_{14}^+=\left\{\begin{array}{rl}
         -\tfrac{ieE_0 f_{-\tau}(\phi)}{(2\hbar\Omega)^2}\Bigg[\tfrac{\gamma\gamma'  \hbar\Omega t_{vc}}{\left(\varepsilon_{c,\tau,s}-\varepsilon_{c',\tau,s}-\hbar\Omega \right)\left(\varepsilon_{c',-\tau,s}-\varepsilon_{v,\tau,s}\right)}-\tfrac{\gamma^2 t_{cv}}{\varepsilon_{c,\tau,s}-\varepsilon_{v,\tau,s}} & \mathrm{XX'} \\ +\tfrac{\gamma'{}^2 \hbar\Omega t_{cv}}{\left(\varepsilon_{c,\tau,s}-\varepsilon_{v',-\tau,s}  \right)\left(\varepsilon_{c,\tau,s}-\varepsilon_{c',-\tau,s}-\hbar\Omega \right)}- \tfrac{\gamma'{}^2 t_{cv}}{\varepsilon_{c',-\tau,s}-\varepsilon_{v',-\tau,s}} \Bigg], \\
         0,& \mathrm{MM'}\\
         0,& \mathrm{2H}
    \end{array} \right.,\nonumber\\
    &v_{23}^-=\left\{\begin{array}{rl}
         \tfrac{ieE_0f_{\tau}(\phi)}{(2\hbar\Omega)^2}\Bigg[\tfrac{\gamma\gamma'\hbar\Omega t_{cv}}{\left(\varepsilon_{c,\tau,s}-\varepsilon_{c',-\tau,s}+\hbar\Omega \right)\left(\varepsilon_{c,\tau,s}-\varepsilon_{v',-\tau,s} \right)}+\tfrac{\gamma'{}^2 t_{vc}}{\varepsilon_{c',-\tau,s}-\varepsilon_{v',-\tau,s}}+& \mathrm{XX'}\\\tfrac{\gamma^2 \hbar\Omega t_{vc}}{\left(\varepsilon_{c',-\tau,s}-\varepsilon_{v,\tau,s} \right)\left(\varepsilon_{c,\tau,s}-\varepsilon_{c',-\tau,s}+\hbar\Omega \right)} +\tfrac{\gamma^2t_{vc}}{\varepsilon_{c,\tau,s}-\varepsilon_{v,\tau,s}} \Bigg],  \\
         0,& \mathrm{MM'}\\
         0,& \mathrm{2H}
    \end{array} \right.,\nonumber\\
    &v_{23}^+=\left\{\begin{array}{rl}
         \tfrac{ieE_0 f_{-\tau}(\phi)}{(2\hbar\Omega)^2}\Bigg[\tfrac{\gamma\gamma' \hbar\Omega t_{cv}}{\left(\varepsilon_{v,\tau,s}-\varepsilon_{v',-\tau,s}+\hbar\Omega \right)\left(\varepsilon_{c,\tau,s}-\varepsilon_{v',-\tau,s} \right)}+\tfrac{\gamma^2 t_{vc}}{\left(\varepsilon_{c,\tau,s}-\varepsilon_{v,\tau,s} \right)}& \mathrm{XX'}\\ +\tfrac{\gamma'{}^2 \hbar\Omega t_{vc}}{\left(\varepsilon_{c',-\tau,s}-\varepsilon_{v,\tau,s} \right)\left(\varepsilon_{v,\tau,s}-\varepsilon_{v',-\tau,s}+\hbar\Omega \right)}+\tfrac{\gamma'{}^2 t_{vc}}{\varepsilon_{c',-\tau,s}-\varepsilon_{v',-\tau,s}} \Bigg], \\
         0,& \mathrm{MM'}\\
         0,& \mathrm{2H}
    \end{array} \right.,\nonumber\\
    &s_{34}=\left\{\begin{array}{rl}
         0,& \mathrm{XX'}  \\
         0,& \mathrm{MM'}\\
         t_{vv},& \mathrm{2H}
    \end{array} \right.,\quad v_{34}=\left\{\begin{array}{rl}
         -\tfrac{\gamma t_{cv}}{2}\left(\tfrac{1}{\varepsilon_{c,\tau,s}-\varepsilon_{v,\tau,s}}+\tfrac{1}{\varepsilon_{c,\tau,s}-\varepsilon_{v',-\tau,s}} \right)& \mathrm{XX'}\\-\tfrac{\gamma't_{vc}}{2}\left(\tfrac{1}{\varepsilon_{c',-\tau,s}-\varepsilon_{v,\tau,s}}  + \tfrac{1}{\varepsilon_{c',-\tau,s}-\varepsilon_{v',-\tau,s}} \right),  \\
         0,& \mathrm{MM'}\\
         0,& \mathrm{2H}
    \end{array} \right.,\nonumber\\
    &w_{34}=\left\{\begin{array}{rl}
         0,& \mathrm{XX'}  \\
         0,& \mathrm{MM'}\\
         -\tfrac{t_{vv}}{2}\left[\tfrac{\gamma^2}{\left(\varepsilon_{c,\tau,s}-\varepsilon_{v,\tau,s} \right)\left(\varepsilon_{c,\tau,s}-\varepsilon_{v',-\tau,s} \right)} + \tfrac{\gamma'{}^2}{\left(\varepsilon_{c',-\tau,s}-\varepsilon_{v,\tau,s} \right)\left(\varepsilon_{c',-\tau,s}-\varepsilon_{v',-\tau,s} \right)} \right],& \mathrm{2H}
    \end{array} \right.,\nonumber\\
    &w_{34}^+=\left\{\begin{array}{rl}
         0,& \mathrm{XX'}  \\
         \tfrac{\gamma\gamma't_{cc}}{2}\left[\tfrac{1}{\left(\varepsilon_{c,\tau,s}-\varepsilon_{v,\tau,s} \right)\left(\varepsilon_{c',-\tau,s}-\varepsilon_{v,\tau,s} \right)} + \tfrac{1}{\left(\varepsilon_{c,\tau,s}-\varepsilon_{v',-\tau,s} \right)\left(\varepsilon_{c',-\tau,s}-\varepsilon_{v',-\tau,s} \right)} \right],& \mathrm{MM'}\\
         0,& \mathrm{2H}
    \end{array} \right.,
\end{align}

\section{Model parameters for WSe${}_2$/MoSe${}_2$}\label{app:WSe2MoSe2parameters}
The model parameters reported in the caption of Fig.\ \ref{fig:PhaseDiagramP} were chosen as follows: the $\kk\cdot\pp$ parameters $\gamma$ (for WSe${}_2$) and $\gamma'$ (for MoSe${}_2$) were taken from fittings to density functional theory + GW results reported by Korm\'anyos et al.\ in Ref.\ \onlinecite{kormanyos2015k}. For the intralayer band gaps $\varepsilon_g$ (for WSe${}_2$) and $\varepsilon_g'$ (for MoSe${}_2$), we used intralayer exciton energies for the corresponding layers in a WSe${}_2$/MoSe${}_2$ hetero-bilayer on top of a SiO${}_2$ substrate, reported by Nagler et al.\ in Ref.\ \onlinecite{Nagler_2017}. For the heterostructure band gap $\Delta_0$, we used the interlayer exciton energy for the same structure, averaging two values reported in the literature: one by Nagler et al.\ in Ref.\ \onlinecite{Nagler_2017} and another by Rivera et al.\ in Ref.\ \onlinecite{Rivera_2015}. The MoSe${}_2$ conduction and valence spin-orbit splittings, $\Delta_c=37\,{\rm meV}$ and $\Delta_v=220\,{\rm meV}$, were obtained from magneto-transport experiments in Ref.\ \onlinecite{Larentis}, and ARPES measurements reported in Ref.\ \onlinecite{Nguyen2019}, respectively.

The tunnelling matrix elements were merely chosen to be the right order of magnitude, according to experimental reports on TMD hetero-bilayers. For instance, in Ref.\ \onlinecite{hX_Nature} Alexeev et al.\ estimate $t_{cc}=26\,{\rm meV}$ for $\theta=1.8^\circ$ P-type MoSe${}_2$/WSe${}_2$ from A exciton hybridization. Unpublished work by one of us finds $t_{vv}=66\,{\rm meV}$ from B exciton hybridization in the same $\theta=1.8^\circ$ sample studied in Ref.\ \onlinecite{hX_Nature}.  Based on this, we have taken a simple hyerarchy of tunnelling energies $t_{vv}=4t_{cc}=100\,{\rm meV}$.

On the other hand, to our knowledge there are no experimental estimates of $t_{cv}$ or $t_{vc}$. This is to be expected, since these tunnelling energies couple states separated by the $\gtrsim 1\,{\rm eV}$ heterostructure band gap, becoming negligible in ordinary circumstances, though important when the gap is closed, as in the present work. Therefore, we have simply chosen $t_{cv}$ and $t_{vc}$ to be of the same order of magnitude as $t_{cc}$.

Finally, the commensurate heterostructure lattice constant $a=3.289\,\text{\AA}$ was estimated as the average of two measured values for bulk WSe${}_2$ and two for MoSe${}_2$, reported in Refs.\ \cite{LatticeConstant1} and \cite{LatticeConstant2}.

\section{Further results for \APH\,structures}

This section presents additional results for 2H-WSe${}_2$/MoSe${}_2$, beyond what is shown in Fig.\ \ref{fig:PhaseDiagramAP}. Figure \ref{fig:2Hgap} shows the size of the topological band gap at the $K$ valley, considering both spin-up and down bands, as a function of the light field parameters. A maximum value of $\Delta_{\rm T}=20\,{\rm meV}$ is found around $(\hbar\Omega,\,eE_0)=(1.92\,{\rm eV},\,152\,{\rm meV/\AA})$.

Figure \ref{fig:2HLeft} shows the topological phase diagram obtained using left circularly polarized light (LCPL), as well as band structures corresponding to a zig-zag nanoribbon of width $L=1500a$. The latter show two pairs of in-gap edge states per spin in the $K'$ sector, but none for the $K$ valley.

Figure \ref{fig:2HLDOS} shows the local density of states (LDOS) at the top and bottom edges of the same zig-zag nanoribbon under right circularly polarized light (RCPL), for different values of the light frequency $\Omega$ at fixed $eE_0=156\,{\rm meV/\AA}$. The group velocities of the in-gap $K$-valley states at the top and bottom edges indicate unidirectional rightward transport on the top, and leftward transport on the bottom edge, corresponding to negative (clockwise) chirality. This is the case for both spins. The spin-down case shows another two pairs of chiral edge bands with the same orientation below the topological gap. Since these overlap with the bulk band $n=2$, they are not visible in the band structures of Fig.\ \ref{fig:PhaseDiagramAP}.

Figure \ref{fig:2Hmorebands} shows nanoribbon band structures for the same case as in Fig.\ \ref{fig:PhaseDiagramAP}, but for multiple values of the field strength $eE_0$ ranging from $50$ to $200\,{\rm meV/\text{\AA}}$. The results show that, although the topological gap in valley $K$ seems to shift toward the trivial gap at valley $K'$, the latter closes, preventing a global gap from forming.

\begin{figure}[t!]
\centering
\includegraphics{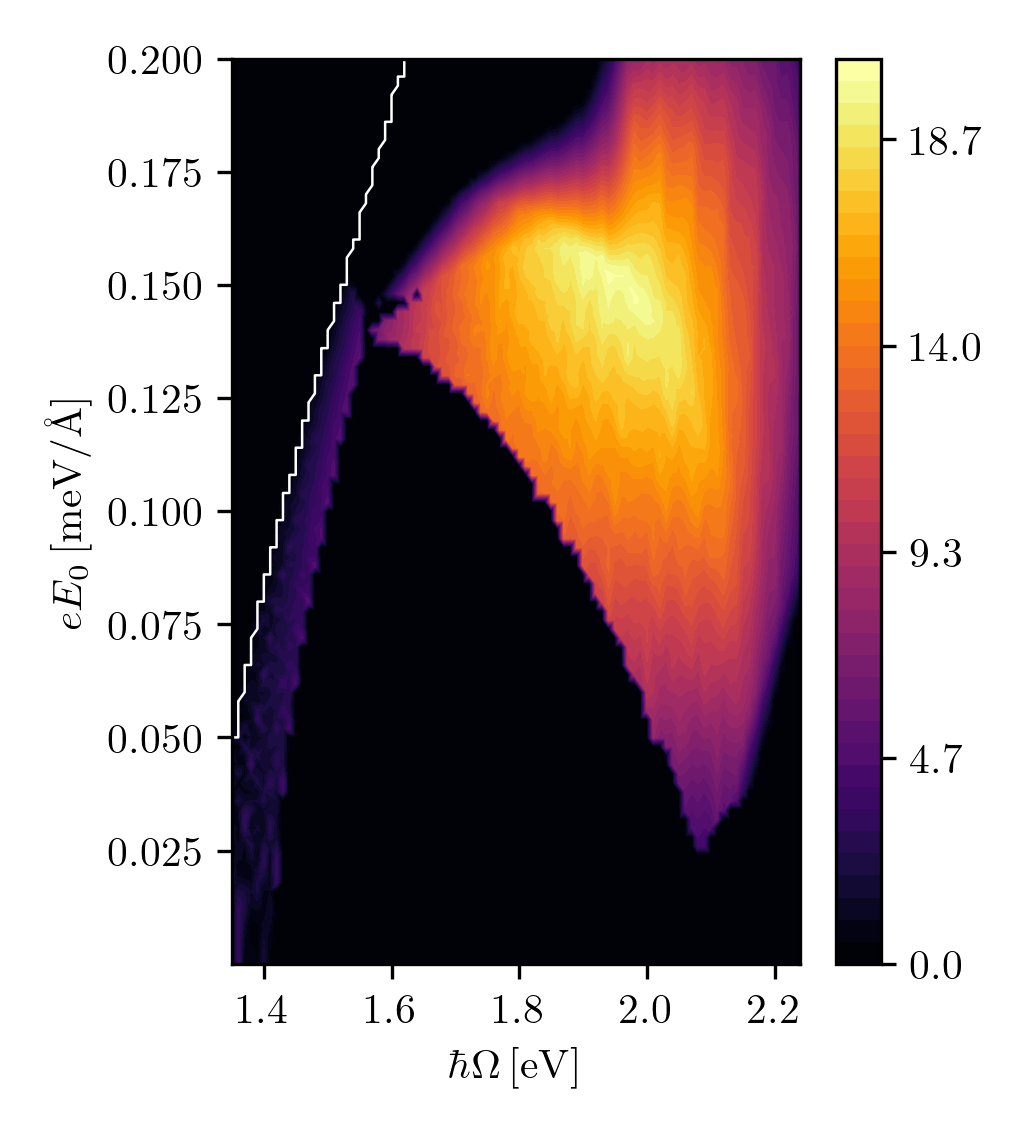}
\caption{Topological gap $\Delta_{\rm T}$ between bands $n=2$ and $n=3$ for 2H-WSe${}_2$/MoSe${}_2$ irradiated with RCPL. The transition line between the trivial (left) and topological (right) phases is shown in white.}\label{fig:2Hgap}    
\end{figure}

\begin{figure}[p!]
    \centering
    \includegraphics[width=\columnwidth]{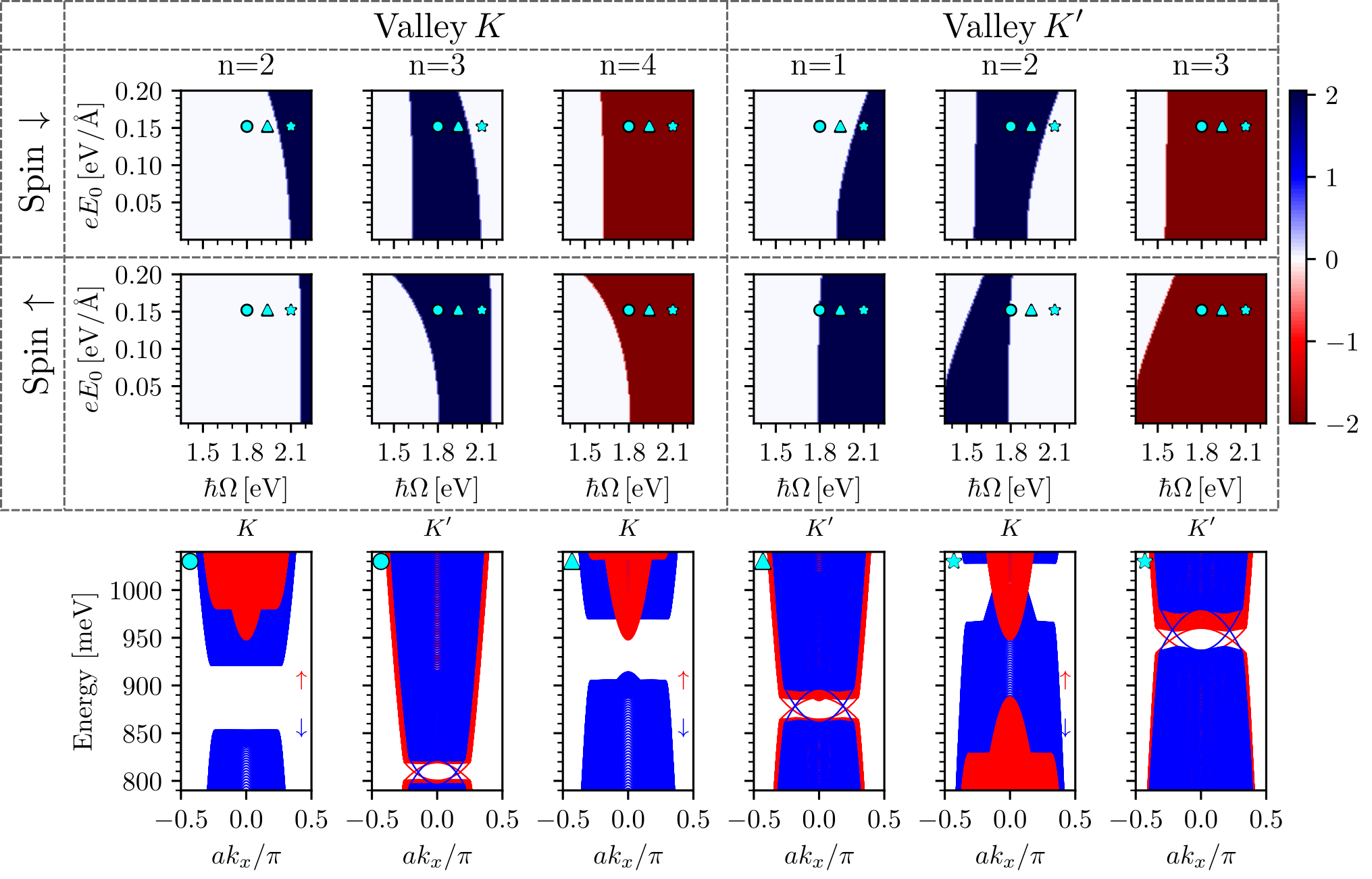}
    \caption{(Top) Topological phase diagram for 2H-WSe${}_2$/MoSe${}_2$ irradiated by LCPL, for both spins at both valleys. Bands with zero Chern number throughout the parameter space explored have been omitted. (Bottom) Band structures for a zig-zag nanoribbon ($L=1500a$) at the points of parameter space indicated with \textcolor{cyan}{$\bullet,\,\blacktriangle$} and \textcolor{cyan}{$\bigstar$} in the phase diagrams, corresponding to light-field parameters $\hbar\Omega=1.80,\,1.92$ and $2.1\,{\rm eV}$ for fixed $eE_0=156\,{\rm meV/\text{\AA}}$. Results are analogous the the case of RCPL shown in Fig.\ \ref{fig:PhaseDiagramAP}, except that the in-gap chiral edge states now originate in the $K'$ valley and have opposite chirality.}
    \label{fig:2HLeft}
\end{figure}

\begin{figure}[p!]
    \centering
    \includegraphics[width=\columnwidth]{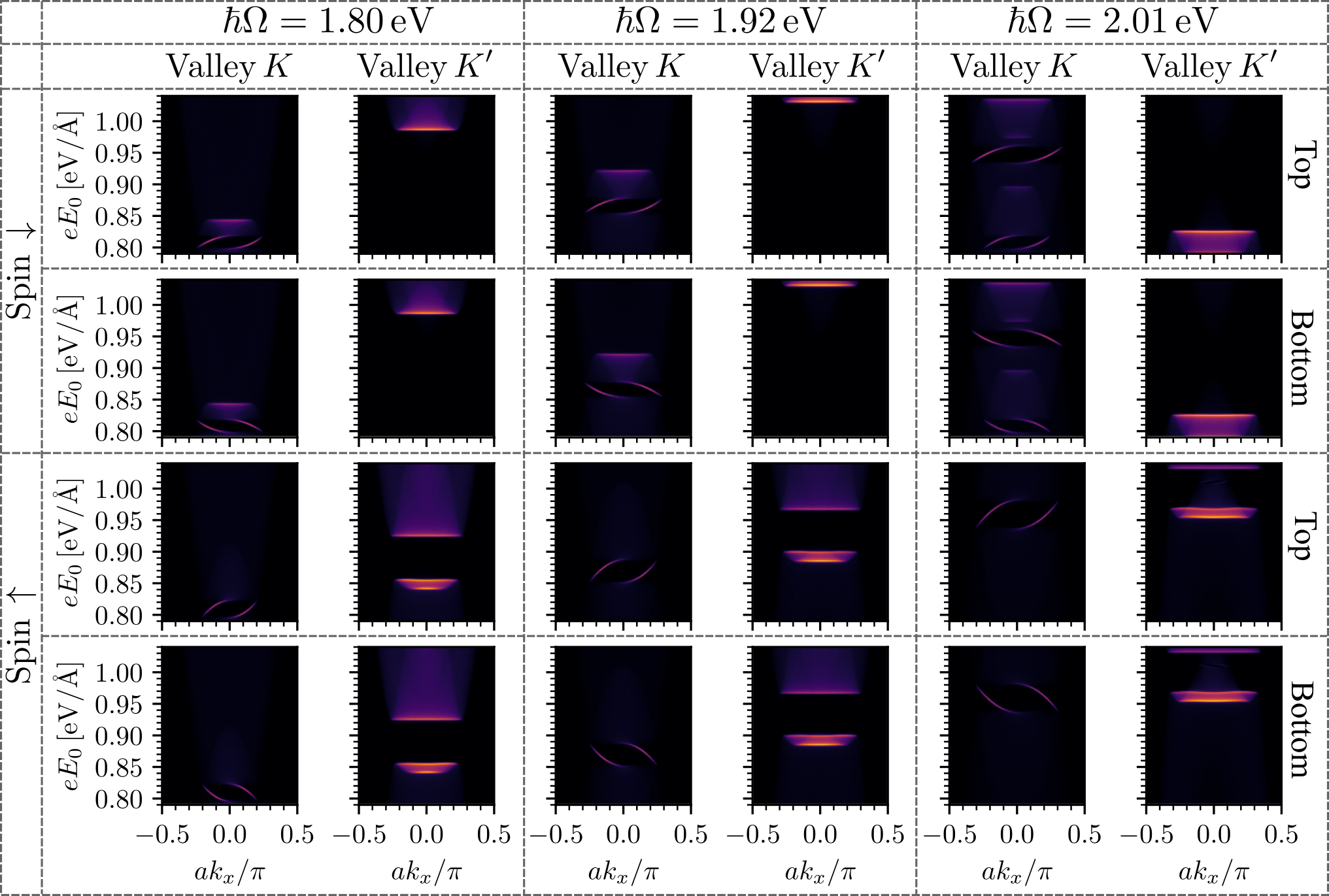}
    \caption{LDOS at the top and bottom edges of a 2H-WSe${}_2$/MoSe${}_2$ zig-zag nanoribbon of width $L=1500a$ under RCPL light. All cases correspond to field intensity $eE_0=156\,{\rm meV}/\text{\AA}$, as in the bottom panels of Fig.\ \ref{fig:PhaseDiagramAP}. Two edge states of negative chirality (i.e., propagating clockwise) per spin appears inside the $K$ valley, and none for the $K'$ valley.}
    \label{fig:2HLDOS}
\end{figure}

\begin{figure}[p!]
    \centering
    \includegraphics[width=\columnwidth]{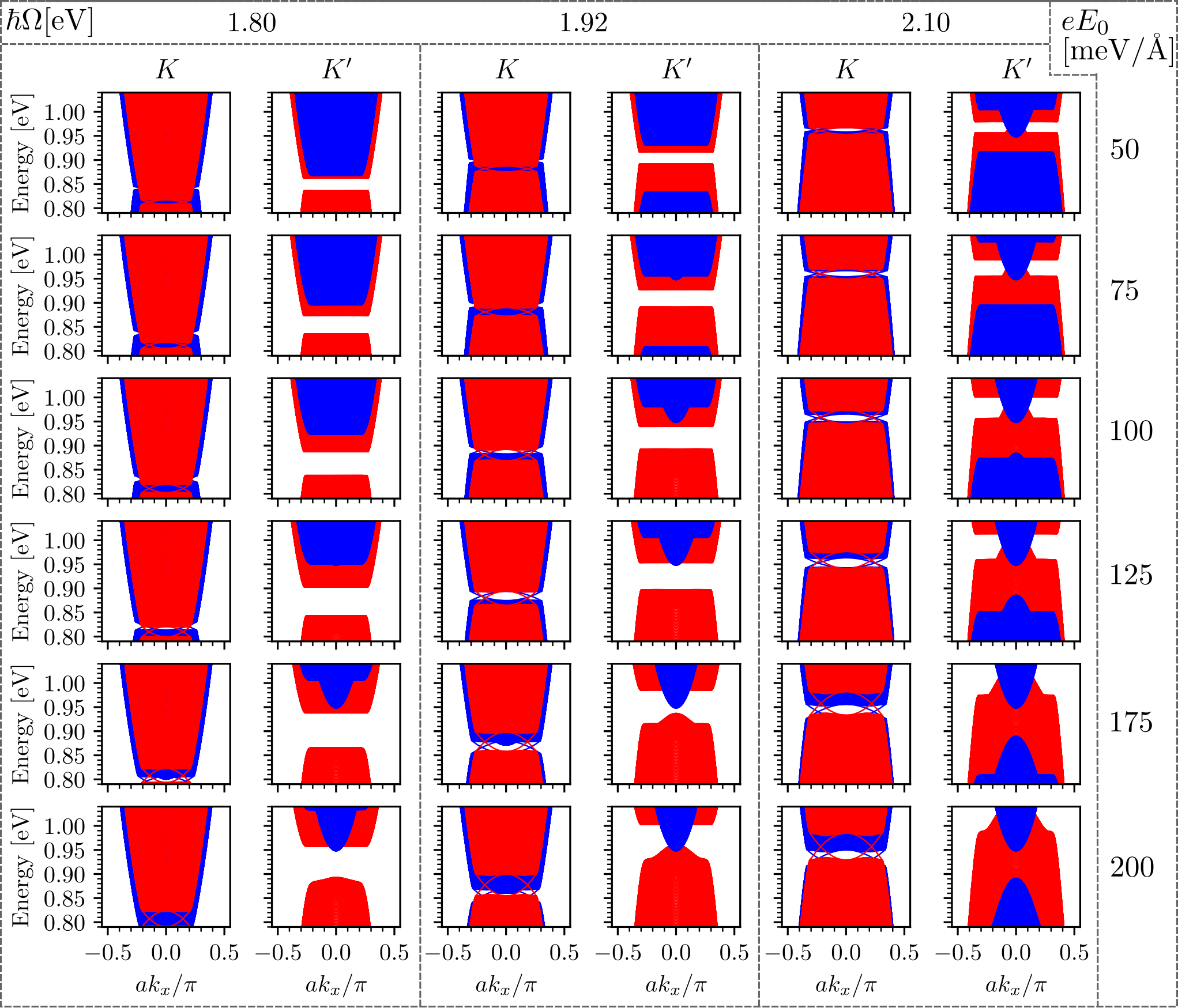}
    \caption{Band structures for a 2H-WSe${}_2$/MoSe${}_2$ zig-zag nanoribbon of width $L=1500a$ under RCPL, for multiple values of the light frequency $\hbar \Omega$ and field strength $eE_0$, complementing the results presented in Fig.\ \ref{fig:PhaseDiagramAP}.}
    \label{fig:2Hmorebands}
\end{figure}
\end{widetext}

\end{document}